\documentclass[amsmath,  floatfix, nofootinbib, prx, aps, usenames, dvipsnames,10pt, allowfontchageintitle, a4paper]{quantumarticle}

\usepackage{algorithm,algpseudocode}

\usepackage{enumitem}

\usepackage{hyperref} 
\usepackage{jabbrv}

\usepackage{enumitem}
\let\olddesc\description
\def\description{\olddesc\setlist[itemize]{leftmargin=*,labelindent=-12pt}}

\usepackage{lineno}

\usepackage{xr}

\usepackage[english]{babel}

\usepackage{graphicx}
\usepackage{caption}
\usepackage{subcaption}

\usepackage{xcolor}

\usepackage{tcolorbox}
\tcbuselibrary{breakable}
\tcbuselibrary{skins}
\tcbset{enhanced, colback=white, arc=0mm, outer arc=0mm, fonttitle=\bfseries, fontupper=\small}
\newtcolorbox[auto counter]{examplebox}[2][]{%
title=Box~\thetcbcounter: #2,#1}

\usepackage{amsmath}
\usepackage{amsthm}
\usepackage{amssymb}
\usepackage{mathtools}
\usepackage{thmtools}

\usepackage{bm}  
\usepackage{bbm}
\usepackage{dsfont}
\usepackage{cases}
\usepackage{nicefrac} 
\usepackage{xfrac}

\usepackage{multicol}

\newtheorem{theorem}{Theorem}
\newtheorem{corollary}[theorem]{Corollary}
\newtheorem{prop}[theorem]{Proposition}
\newtheorem{lemma}[theorem]{Lemma}

\newcommand{\ket}[1]{\left| #1 \right>} 
\newcommand{\bra}[1]{\left< #1 \right|} 

\let\epsilon\relax
\def\epsilon{\varepsilon}
\let\phi\relax
\def\phi{\varphi}

\begin{document}

\title{Measurement-based quantum computation from Clifford quantum cellular automata}

\author{Hendrik Poulsen Nautrup\**\,}
\email{hendrik.poulsen-nautrup@uibk.ac.at}
\affiliation{University of Innsbruck, Institute for Theoretical Physics, Technikerstr. 21a, A-6020 Innsbruck, Austria}

\author{Hans J.~Briegel}
\affiliation{University of Innsbruck, Institute for Theoretical Physics, Technikerstr. 21a, A-6020 Innsbruck, Austria}

\date{}

\begin{abstract}\vspace{-0.5cm}

\noindent
Measurement-based quantum computation (MBQC) is a paradigm for quantum computation where computation is driven by local measurements on a suitably entangled resource state. In this work we show that MBQC is related to a model of quantum computation based on Clifford quantum cellular automata (CQCAs). Specifically, we show that certain MBQCs can be directly constructed from CQCAs which yields a simple and intuitive circuit model representation of MBQC in terms of quantum computation based on CQCA.
We apply this description to construct various MBQC-based Ansätze for parameterized quantum circuits, demonstrating that the different Ansätze may lead to significantly different performances on different learning tasks. In this way, MBQC yields a family of hardware-efficient Ansätze that may be adapted to specific problem settings and are particularly well suited for architectures with translationally invariant gates such as neutral atoms.
\vspace{-2pt}
\end{abstract} 
\maketitle

\section{Introduction}\vspace{-0.2cm}
The aim of this work is to construct measurement-based quantum computation (MBQC) on a lattice from a circuit model quantum computation based on quantum cellular automata (QCAs). In doing so, we reveal an illustrative and pictorial way of understanding MBQC that also gives a new perspective on the intriguing and exciting world of MBQC.
At the same time, this perspective yields a practical way of constructing families of MBQC-based Ansätze for parameterized quantum circuits (PQCs).

\begin{figure*}[ht]\vspace{-0.0cm}
\centering
  \centering
  \includegraphics[width=0.68\textwidth]{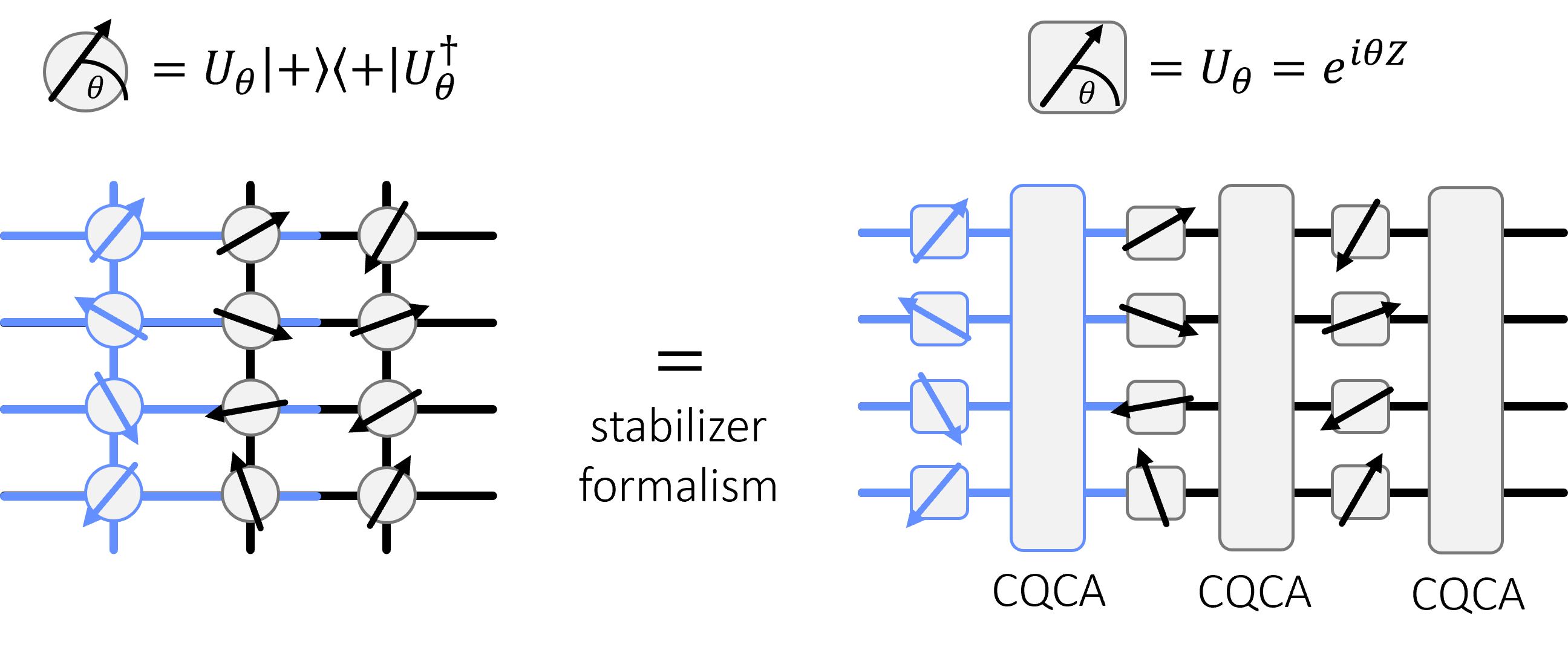} 
\caption{\textbf{Measurement-based quantum computation (MBQC) as a quantum computation based on Clifford quantum cellular automata (CQCAs).} In this work, we will discuss the relation between MBQC (left) where measurements in the $XY$-plane are performed on a suitably entangled resource state and a model of quantum computation based on CQCA (right) where repeated applications of CQCA are interspersed with local $Z$-rotations. This relationship becomes apparent in the Heisenberg picture using the stabilizer formalism.}
\label{fig:intro}
\end{figure*}

MBQC is a model of computation where a unitary is implemented by measurements on a suitably entangled resource state~\cite{raussendorf_oneway_2001,briegel_persistent_2001} (see Fig.~\ref{fig:intro}(left)). While this model has been shown to be universal across various resource states and suitable choices of measurement bases, we will focus on so-called graph states~\cite{hein_entanglement_2005} defined on regular lattices and measurements in the $XY$-plane.

One-dimensional (1D) QCAs are translationally invariant, locality preserving unitaries acting on a 1D chain of qubits. These are unitaries that map local observables (i.e., observables with support on a constant number of qubits) onto local observables. They can be represented as products of constant-depth, local unitaries. A QCA-based quantum computation intersperses the application of QCAs with local rotations (see Fig.~\ref{fig:intro}(right)). In fact, we only consider so-called Clifford QCAs (CQCAs) which map products of Pauli operators to products of Pauli operators and are therefore themselves products of local Clifford operations. As CQCAs are the natural analog to classical cellular automata, we refer to this model of quantum computation as cellular automaton-based quantum computation (CAQC). We will show that CAQC is universal.

We want to understand MBQC and CAQC as implementing a specific unitary. Our final goal is to translate measurements in the MBQC picture to local rotations in the CAQC picture which then enables us to explicitly write the effective unitary implemented through an MBQC. 
Instead of relying on the Schrödinger picture where we understand the action of a unitary in terms of its action on a quantum state, we consider the Heisenberg picture where we understand a unitary as transforming the space of local observables. This picture naturally involves CQCAs and the stabilizer formalism which we will use to understand the basic blocks of MBQC in terms of CAQC.

%
%

As part of this work, we give new perspectives and insights into MBQC through CAQC. 
While the connection between MBQC and QCAs was first established in Ref.~\cite{raussendorf_quantum_2005}, it was not until Refs.~\cite{stephen_subsystem_2019,raussendorf_computationally_2019} that it was properly generalized. There, the authors relied on an efficient CQCA formalism introduced in Ref.~\cite{schlingemann_on_2008} to construct so-called computational phases of matter. Our perspective expands on the work in Ref.~\cite{stephen_subsystem_2019} making it accessible to the field of quantum computation by solely relying on the Heisenberg picture and stabilizer formalism~\cite{gottesmann_stabilizer_1997,nielsen_quantum_2000}.

Aside from these insights, our perspective on MBQC based on CQCA  naturally gives rise to an MBQC-based approach to PQC that may be of use for the design of hardware-efficient and problem specific ansätze. We show that this family of MBQC-based PQCs covers a variety of learning models that perform very differently depending on the learning problem. Specifically, we consider biased quantum datasets for supervised learning where the labels are generated from one model and have to be learned by the others~\cite{huang_power_2021,havlivcek_supervised_2019}.

We start by introducing CAQC in Sec.~\ref{sec:QCAQC} which relies heavily on the circuit model and Heisenberg picture. The essence of a CAQC is captured by Theorem~\ref{theorem:qcaqc_gates}. In Sec.~\ref{sec:stabilizers}, we introduce the stabilizer formalism in so far as to describe measurement-based teleportation in the Heisenberg picture. In Sec.~\ref{sec:mbqc}, we then use the stabilizer formalism to understand that MBQC has a simple representation in terms of CAQC. To this end, we formulate MBQC in terms of Algorithm~\ref{alg:mbqc} which is entirely based on a CQCA. The result of this algorithm is captured by Theorem~\ref{theorem:mbqc_unitary} while Theorem~\ref{theorem:stab_state} connects Algorithm~\ref{alg:mbqc} to a resource state for MBQC. Propositions~\ref{prop:qcaqc_extended} and~\ref{prop:stab_extended} extend this formalism to enable universal CAQC and MBQC for any CQCA. In Sec.~\ref{sec:pqc}, we apply the formalism of CAQC to investigate MBQC-based Ansätze for PQCs.

\section{Cellular automaton-based quantum computation}\label{sec:QCAQC}
In this section, we introduce the concept of Clifford quantum cellular automata (CQCAs) which we use to construct a translationally invariant model for universal quantum computation as shown in Fig.~\ref{fig:intro}(right).

\subsection{Clifford QCA}
In this work, we consider chains of qubits, i.e., qubits arranged equidistantly along a line such that there is a natural notion of locality. In fact, we will assume that qubits are arranged on a ring, i.e., a chain with periodic boundary conditions, but the results in this work extend to chains with boundaries too.
1D Quantum cellular automata (QCAs) are defined as \emph{translationally invariant, locality preserving unitaries} acting on a ring of $N$ qubits. To make sense of this definition, we will understand QCAs through their action on the group of local observables for each qubit $i=1,...,N$ spanned by the local Pauli basis $X_i,Y_i,Z_i$. This is equivalent to saying that we consider the Bloch sphere representation of qubits in the Heisenberg picture where we evolve observables instead of states. The advantage of considering the action of QCAs on the observables rather than states is that locality and translational invariance are particularly well encapsulated within this formalism as we will see in the following.

A 1D QCA is defined by a unitary, also called transition function $T$, such that for any onsite observable $O_i$ acting on vertex $i$, $T(O_i)\equiv TO_iT^\dagger$ is localized within a region $[i-r,i+r]$ for some constant $r\in\mathbb{N}$ independent of $N$ ($T$ is locality preserving) and given $T(O_i)$ for any such observable $O_i$, we also know $T(O_{i'})$ for the observable at any other vertex $i'=1,...,N$ ($T$ is translationally invariant). We see that a QCA is completely specified by its action on the local observables, also called the \emph{local transition rule}. 

In this work, we further restrict the class of QCAs to so-called \emph{Clifford} QCAs (or CQCAs), which are QCAs that map products of Pauli operators to products of Pauli operators. CQCAs are then completely specified by their action $T(X_i),T(Z_i)$ since $T(Y_i)=T(iX_iZ_i)=iT(X_i)T(Z_i)$. In particular, CQCAs can be described by products of constant-depth, local and translationally invariant Clifford circuits. 
Moreover, we only consider CQCA that are symmetrical around any vertex $i$. 
Further note that we will use a product and tensor product interchangeably by assuming that any local observable is trivially extended by identity operators to act on the entire Hilbert space. That is, for example,  $X_1X_2\equiv X_1\otimes X_2\otimes \mathbbm{1}$ for a system size $N=3$.
%
\begin{figure*}[ht!]\vspace{-0.0cm}
\centering
  \centering
  \includegraphics[width=0.99\textwidth]{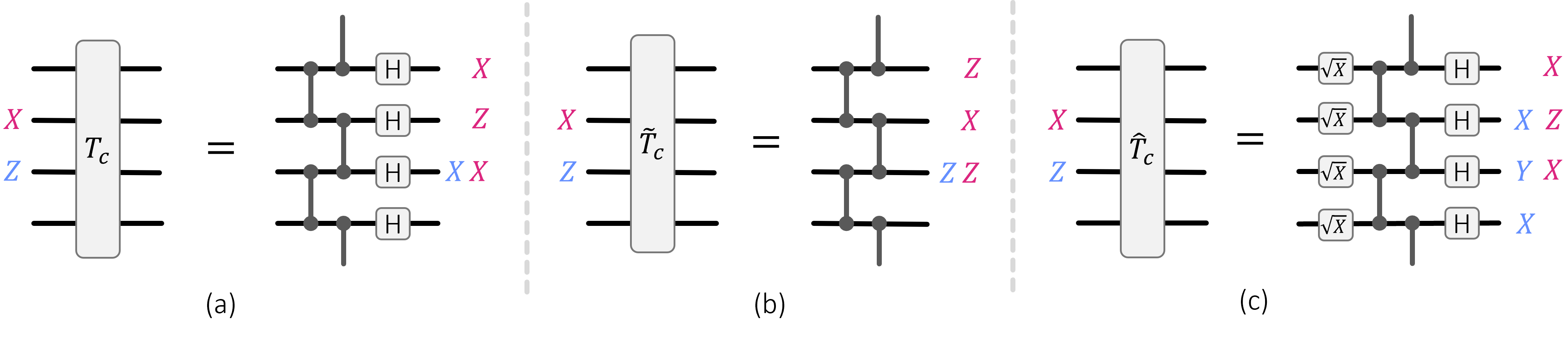} 
\caption{\textbf{Example CQCAs and their action on the basis of observables.} CQCAs are translationally invariant, locality preserving unitaries acting on a ring of $N$ qubits that map products of Pauli operators to products of Pauli operators. They are defined by their action on $X_i,Z_i$ for any one qubit $i\in\{1,...,N\}$. (a) Example of the cluster CQCA $T_c$ which is a so-called \emph{glider} CQCA. (b) Example of a \emph{periodic} CQCA $\tilde{T}_c$ that has a fixed period $L=2$ for any $N$. (c) Example of a \emph{fractal} CQCA $\hat{T}_c$ that is neither a glider nor a periodic CQCA. Here $H$ is a Hadamard gate, lines connecting qubits represent $CZ$-gates and $\sqrt{X}=HSH$ where $S=\sqrt{Z}$ is a phase gate.}
\label{fig:cqca_examples}
\end{figure*}
%
Let us now consider some relevant examples of CQCAs depicted in Fig.~\ref{fig:cqca_examples}. The first example in Fig.~\ref{fig:cqca_examples}(a) is the so-called \emph{cluster state CQCA} (or cluster CQCA for short) $T_c$ which is defined as,
\begin{align}
    T_c(X_i)&=X_{i-1}Z_iX_{i+1}\nonumber\\
    T_c(Z_i)&=X_i\label{eq:qca_cluster}.
\end{align}
We see that this is a valid QCA as it is a unitary that is translationally invariant, i.e., it acts the same way on each qubit $i=1,...,N$ and it is locality preserving as $T(O_i)$ acts within a region $[i-1,i+1]$ for any onsite observable $O_i$. We further find by the relations in Eq.~\eqref{eq:qca_cluster} that it is a CQCA.
$T_c$ is a so-called \emph{glider} CQCA because there exist a local Pauli observable 
on which $T_c$ acts as a translation. 
As, we can see from Eq.~\eqref{eq:qca_cluster} and Fig.~\ref{fig:cqca_examples}(a), $T(Z_{i-1}X_{i})=Z_iX_{i+1}$.

As a second example, consider the CQCA in Fig.~\ref{fig:cqca_examples}(b) which acts as 
\begin{align}
    \tilde{T}_c(X_i)&=Z_{i-1}X_iZ_{i+1}\nonumber\\
    \tilde{T}_c(Z_i)&=Z_i.\label{eq:qca_periodic}
\end{align}
This is a so-called \emph{periodic} CQCA because there exists a constant $p$ independent of the system size $N$ such that $\tilde{T}_c^p$ acts as the identity, i.e., $\tilde{T}_c^p=id$. As we can see from the local transition rules $\tilde{T}_c^2=id$.

As the last example, we investigate the CQCA in Fig.~\ref{fig:cqca_examples}(c) which is described by,
\begin{align}
    \hat{T}_c(X_i)&=X_{i-1}Z_iX_{i+1}\nonumber\\
    \hat{T}_c(Z_i)&=X_{i-1}Y_iX_{i+1}.\label{eq:qca_fractal}
\end{align}
This CQCA is neither a glider nor a periodic CQCA and therefore, a so-called \emph{fractal} CQCA. 
These three classes have been first described and analyzsed in Ref.~\cite{schlingemann_on_2008}.

At this point, let us state a Lemma which is at the heart of universality of CQCA-based quantum computing~\cite{stephen_subsystem_2019}:
\begin{lemma}\label{lemma:period}
For any glider and periodic CQCA $T$ defined on a ring of size $N$, there exists $L=\mathcal{O}(N)$ and $L=\mathcal{O}(1)$, respectively, such that $T^L=id$.\\
For any fractal CQCA $T$ defined on a ring of size $N=2^k$ with $k\in\mathbb{N}$, there exists $L=\mathcal{O}(N)$ such that $T^L=id$.
\end{lemma} 
That is, any CQCA has a period that scales at most linear with the system size $N$.

For a more thorough and in-depth introduction to QCAs and CQCAs in particular, we refer to Refs.~\cite{schlingemann_on_2008,stephen_subsystem_2019}. Notably, there the authors introduce a classically efficient description of CQCA in terms of $2\times 2$ matrices over Laurent polynomials.

\subsection{Simple and entangling CQCAs}
Let us further restrict the class of CQCAs as this will simplify the discussion of universality in the following subsection. 

We call a CQCA \emph{simple} if,
\begin{align}
    T(Z_i)&=X_i\nonumber\\
    T(X_i)&=Z_i\times P_X,\label{eq:cqca_simple}
\end{align}
where $P_X$ is a product of Pauli-$X$ operators in a constant-size region around (and including) vertex $i$ such that $T$ is a CQCA as discussed in the previous subsection. $T_c$ in Eq.~\eqref{eq:qca_cluster} is an example of a simple CQCAs, but $\tilde{T}_c$ in Eq.~\eqref{eq:qca_periodic} and $\hat{T}_c$ in Eq.~\eqref{eq:qca_fractal} are not simple.

We call a CQCA \emph{entangling} if,
\begin{align}
    w(T(Z_i))>1 \hspace{0.2cm}\textrm{ or }\hspace{0.2cm} w(T(X_i))>1,\label{eq:cqca_entangling}
\end{align}
where $w(P)$ counts the number of non-identity terms in a tensor product of Pauli operators $P$ and is often referred to as the \emph{weight}. All CQCAs $T_c,\tilde{T}_c$ and $\hat{T}_c$ are entangling by this definition as $w(T_c(X_i))=w(\tilde{T}_c(X_i))=w(\hat{T}_c(X_i))=3$.

While the restriction to entangling CQCAs is necessary to prove universality of CAQC and MBQC in Sec.~\ref{sec:mbqc}, we will show how to circumvent the restriction to simple CQCAs in Sec.~\ref{sec:nonsimple}.
%
\begin{figure}[ht!]\vspace{-0.0cm}
\centering
  \centering
  \includegraphics[width=\linewidth]{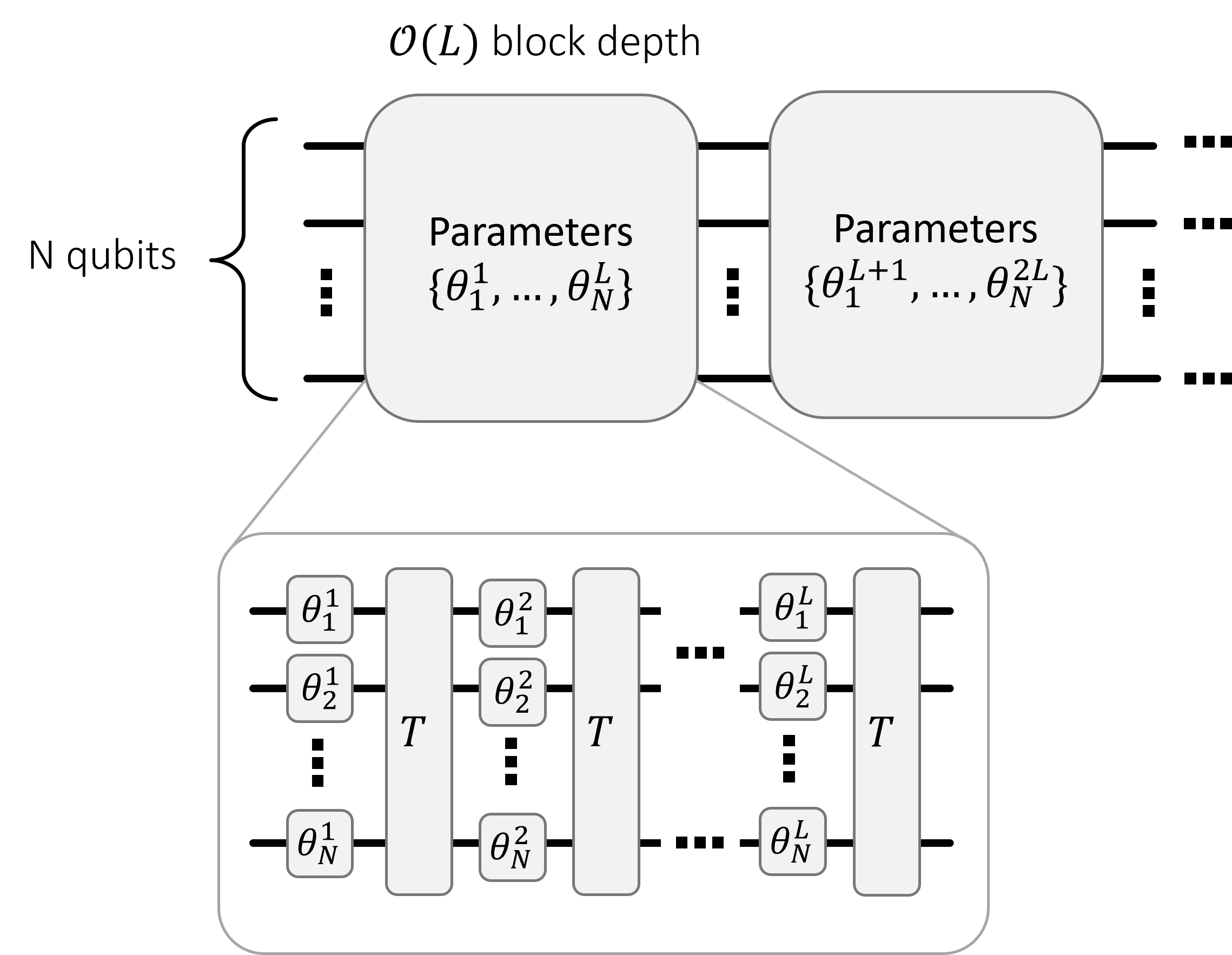} 
\caption{\textbf{Cellular automaton-based quantum computation (CAQC) is constructed from blocks.} The block depth of a CAQC is defined by the period $L$ of the corresponding CQCA $T$ such that each block acts as $T^L=id$ if no rotations are present. Each block has $L\times N$ independent parameters, e.g., $\{\theta^j_i\}_{i,j}$ for $j=1,...,L$, $i=1,...,N$ in the first block, defining the rotation angles for $\exp(i\theta^j_i Z_i)$ acting on qubit $i$ at depth $j$. As we will see, each block yields a universal set of gates, parameterized by $\{\theta^j_i\}_{i,j}$, such that their concatenation enables universal quantum computation.}
\label{fig:qcaqc_blocks}
\end{figure}
%
\subsection{Universal CAQC}
In this section we construct universal quantum circuits from repeated applications of CQCA $T$ and local rotations, which we call CQCA-based quantum computing or CAQC (which is formalized in Theorem~\ref{theorem:qcaqc_gates}). To this end, consider a circuit model computation on $N$ qubits separated into blocks of depth $\mathcal{O}(L)$ (see Fig.~\ref{fig:qcaqc_blocks}) where $L$ is the period of the $T$, i.e., $T^L=id$.

As depicted in Fig.~\ref{fig:qcaqc_blocks}, each block consists of $L$ applications of $T$ interspersed with $N$ $Z$-rotations $e^{i\theta Z}$ before every $T$. Therefore, each block comprises $N\times L$ independent parameters. We will show, by explicit construction, that each block can implement a universal set of gates if $T$ is simple and entangling.

CQCAs are fully described by their actions on the local Pauli observables $X_i,Z_i$. The latter also describes the special unitary group $SU(2)$ of local rotations through the exponential map $e^{\theta P}\in SU(2)$ for $P\in\mathfrak{su}(2)$ (where $iX_i,iZ_i,iY_i$ are the generators of the Lie algebra $\mathfrak{su}(2)$). Through the local rules of CQCAs, we can then specify the unitary implemented by a $Z$-rotation $R_Z(2\theta^{j}_i)=\exp(i\theta^{j}_i Z_i)$ acting on qubit $i$ at depth $j$ within one block:
\begin{align}
    T^{L-j+1}e^{i\theta^{j}_i Z_i}T^{j-1}=e^{i\theta^{j}_i T^{L-j+1}(Z_i)},\label{eq:qcaqc_rot}
\end{align}
where we used $T^L=id$ and the commutation relation $TZ=TZT^\dagger T= T(Z)T$, defining $T(Z):=TZT^\dagger$. That is, the rotations that can be achieved within each block are of the form $\exp(i\theta T^k(Z_i))$ for $k=1,...,L$. 
In particular, given a simple CQCA as defined in Eq.~\eqref{eq:cqca_simple}, we can implement any one-qubit rotation as,
\begin{align*}
    e^{i\theta T(Z_i)}&=e^{i\theta X_i}\\
    e^{i\theta T^L(Z_i)}&=e^{i\theta Z_i},
\end{align*}
where we only used in the first equation that $T$ is simple. 
Because we also assume $T$  to be entangling as defined by Eq.~\eqref{eq:cqca_entangling}, we can construct an entangling gate as follows,
\begin{align*}
    e^{i\theta T^2(Z_i)}=e^{i\theta Z_i\times P_X},
\end{align*}
where $P_X$ acts on at least two more qubits within the region $[i-r,i+r]$ for some constant $r$ because it is entangling. Therefore, we have shown that each parameterized block as shown in Fig.~\ref{fig:qcaqc_blocks}, can implement a universal set of gates if $T$ is simple and entangling. This means that the concatenation of blocks yields a universal quantum computation up to an $\mathcal{O}(N)$ overhead because $L=\mathcal{O}(N)$ by Lemma~\ref{lemma:period}. Note however, that the overhead only appears because we consider a limited set of gates $\exp(i\theta T^k(Z_i))$ for $k=1,2,L$ within one block ignoring $k=3,4,...,L-1$.
In Sec.~\ref{sec:nonsimple}, we will discuss how to extend a CAQC (and the corresponding MBQC) to yield universality for any entangling CQCA, even those that are not simple. 

%
\begin{figure}[b!]\vspace{-0.0cm}
\centering
  \centering
  \includegraphics[width=\linewidth]{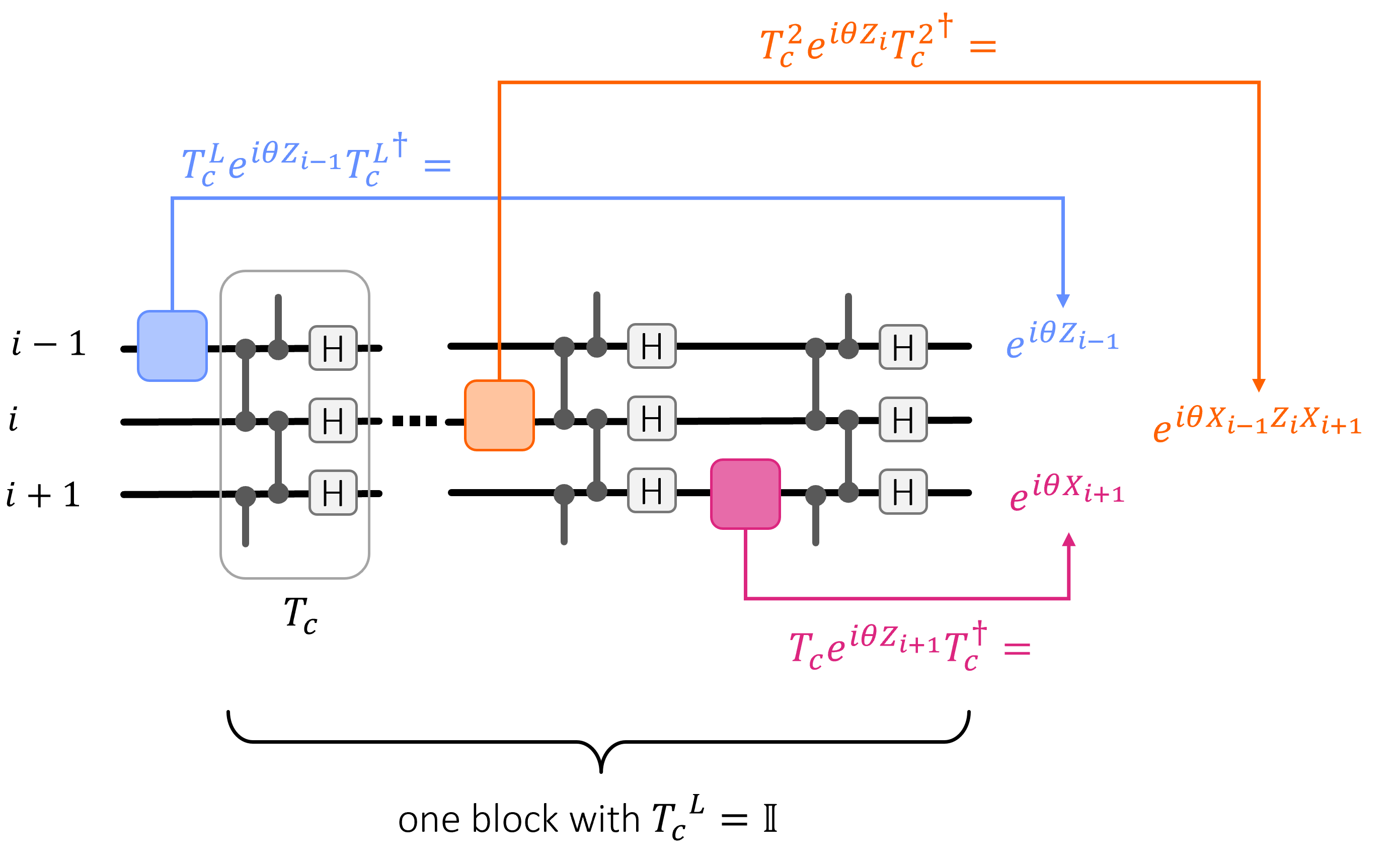} 
\caption{\textbf{Constructing a universal gate set for one block of CAQC based on $T_c$.} As we consider periodic boundary conditions and CQCAs are translationally invariant, it is sufficient to show that within a constant-size region of qubits (here: qubits $i-1,i,i+1$), we can construct any single-qubit rotation and one entangling gate. Depicted in blue on qubit $i-1$, we show that a $Z$-rotation before the full block is equivalent to a $Z$-rotation at the end of the block as $T_c^L=id$. Depicted in orange on qubit $i$ we show that a $Z$-rotation at depth $L-1$ within the block (where we count depth by the layers of $Z$-rotations), yields a 3-qubit entangling gate as $T_c^2(Z_i)=X_{i-1}Z_iX_{i+1}$. Depicted in magenta on qubit $i+1$ we show that a $Z$-rotation at depth $L$ within the block, yields a $X$-rotation as $T(Z_{i+1})=X_{i+1}$.}
\label{fig:qcaqc_cluster}
\end{figure}
%
As an example, we consider in Fig.~\ref{fig:qcaqc_cluster} a set of universal gates within one block for the CQCA $T_c$ defined in Eq.~\eqref{eq:qca_cluster}. As described above, we find,
\begin{align*}
    e^{i\theta T_c(Z_i)}&=e^{i\theta X_i},\\
    e^{i\theta T_c^L(Z_i)}&=e^{i\theta Z_i},\\
    e^{i\theta T_c^2(Z_i)}&=e^{i\theta X_{i-1}Z_iX_{i+1}},
\end{align*}
which are universal for quantum computation in the circuit picture~\cite{stephen_subsystem_2019}. 

To conclude let us use our former insights to formalize the unitary implemented by each block of CAQC.
\newpage
\begin{theorem}\label{theorem:qcaqc_gates}
Consider a CQCA $T$ with a period $L$ acting on a ring of $N$ qubits. Let a block of CQCA-based quantum computation consist of $L$ sequential applications of $T$, where the $j^{\mathrm{th}}$ application of $T$ is preceded by $N$ $Z$-rotations $\exp(i\theta^j_iZ_i)$ acting on qubits $i=1,...,N$ with parameters $\{\theta_1^j,...,\theta_N^j\}$ for all $j=1,...,L$ (see Fig.~\ref{fig:qcaqc_blocks}). This block implements the following unitary:
\begin{align}\label{eq:qcaqca_gates}
    \prod_{j=L,...,1}\prod_{i=N,...,1}\exp\left({i\theta^{j}_iT^{L-j+1}(Z_i)}\right).
\end{align}
\end{theorem}
Note that in Eq.~\eqref{eq:qcaqca_gates} the order of terms in the product is read from right to left. That is, the rightmost term in the product $\exp(i\theta^1_1Z_1)$ is the term that is applied first to the state while the leftmost term $\exp(i\theta^L_N X_N)$ is the last to be applied.

\section{Stabilizer formalism}\label{sec:stabilizers}

In this section, we will revisit the stabilizer formalism so that we can discuss in Sec.~\ref{sec:mbqc} how the state space in an MBQC is transformed through repeated measurements.

\subsection{Stabilizer states}
Quantum states can be described by their Bloch sphere representation using the Pauli operators as a basis of the single qubit space of observables (i.e., $2\times2$ Hermitian matrices). Using the tensor product, this formalism can, in principle, be extended to $N$-qubit states (i.e., $2^N\times2^N$ Hermitian matrices) with potentially exponentially many coefficients describing the correlations between qubits.
While a general quantum state does not allow an efficient classical description in this way, some special states do. One class of such states are \emph{stabilizer states}. These are states that can be described by a set $\mathcal{S}\subset\mathcal{P}_N$ generated by $N$ commuting operators $\{S_i\}_{i=1,...,N}$ under multiplication (denoted as $\mathcal{S}=\langle S_i\rangle_{i=1,...,N}$) where each $S_i$ cannot be written as the product of any other generators (we say, all $S_i$ are \emph{independent}) and $-\mathbbm{I}\notin\mathcal{S}$. Here, we denote by $\mathcal{P}_N$ the Pauli group, i.e., the set of all products of Pauli operators on $N$ qubits. Then, the corresponding stabilizer state $\ket{\mathcal{S}}$ is defined as the $+1$ eigenstate of all generators $S_i\in\mathcal{S}$ such that,
\begin{align}
    \ket{\mathcal{S}}\bra{\mathcal{S}}=\frac{1}{|\mathcal{S}|}\sum_{S\in\mathcal{S}}S.\label{eq:stab_proj}
\end{align}
That is, the state is fully described by the generators $S_i$ of $\mathcal{S}$. We call $\mathcal{S}$ the stabilizer group and its elements stabilizers.

As a first example, consider the two eigenstates of two $X$-operator $\ket{++}=\frac{1}{\sqrt{2}}(\ket{0}_1+\ket{1}_1)\otimes\frac{1}{\sqrt{2}}(\ket{0}_2+\ket{1}_2)$. As $X\ket{+}=\ket{+}$, there are two independent (under multiplication) generators $X_i$ for $i=1,2$ such that $X_i\ket{++}=\ket{++}$. Here, we write $X_i$ to indicate Pauli-$X$ acting on qubit $i$ and identity elsewhere, i.e., $X_1\equiv X_1\otimes\mathbbm{1}$, $X_2\equiv \mathbbm{1}\otimes X_2$ and $X_1X_2\equiv X_1\otimes X_2$. For this example, we can easily verify,
\begin{align*}
    \ket{++}\bra{++}=\frac{1}{4}(\mathbbm{I}+X_1+X_2+X_1X_2).
\end{align*}
Instead of writing down the state explicitly as we did above, we can describe the $\ket{+}$-state for any number of qubits, say $\ket{+}^{\otimes N}$, by their stabilizer group as follows:
\begin{align}
    \mathcal{S}_X=\left\langle X_i\right\rangle_{i=1,...,N}.\label{eq:x-state}
\end{align}

As a second example, consider the Bell state $\ket{\Psi^+}=\frac{1}{\sqrt{2}}(\ket{00}+\ket{11})$. There are two independent stabilizers $X_1X_2$ and $Z_1Z_2$ such that $X_1X_2\ket{\Psi^+}=Z_1Z_2\ket{\Psi^+}=\ket{\Psi^+}$. We find,
\begin{align*}
    \ket{\Psi^+}\bra{\Psi^+}=\frac{1}{4}(\mathbbm{I}+X_1X_2+Z_1Z_2-Y_1Y_2)
\end{align*}
because $XZ=iY$. Equivalently, the corresponding stabilizer group that defines this state exactly is,
\begin{align*}
    \mathcal{S}_{\Psi^+}=\langle X_1X_2, Z_1Z_2\rangle.
\end{align*}

As the last and most relevant example, we consider the family of so-called graph states $\{\ket{G}\}_G$ which are defined for any graph $G=(V,E)$ as
\begin{align}
    \ket{G}=\prod_{\{ i,j\}\in E}CZ_{i,j}\ket{+}^{\otimes N},\label{eq:graph_state}
\end{align}
where each of the $N=|V|$ vertices $v\in V$ is associated with one qubit and (undirected) edges $\{ v_1,v_2\}\in E\subset V\times V$ are unordered pairs of connected vertices.

\begin{figure}[t!]\vspace{-0.0cm}
\centering
  \centering
  \includegraphics[width=0.6\linewidth]{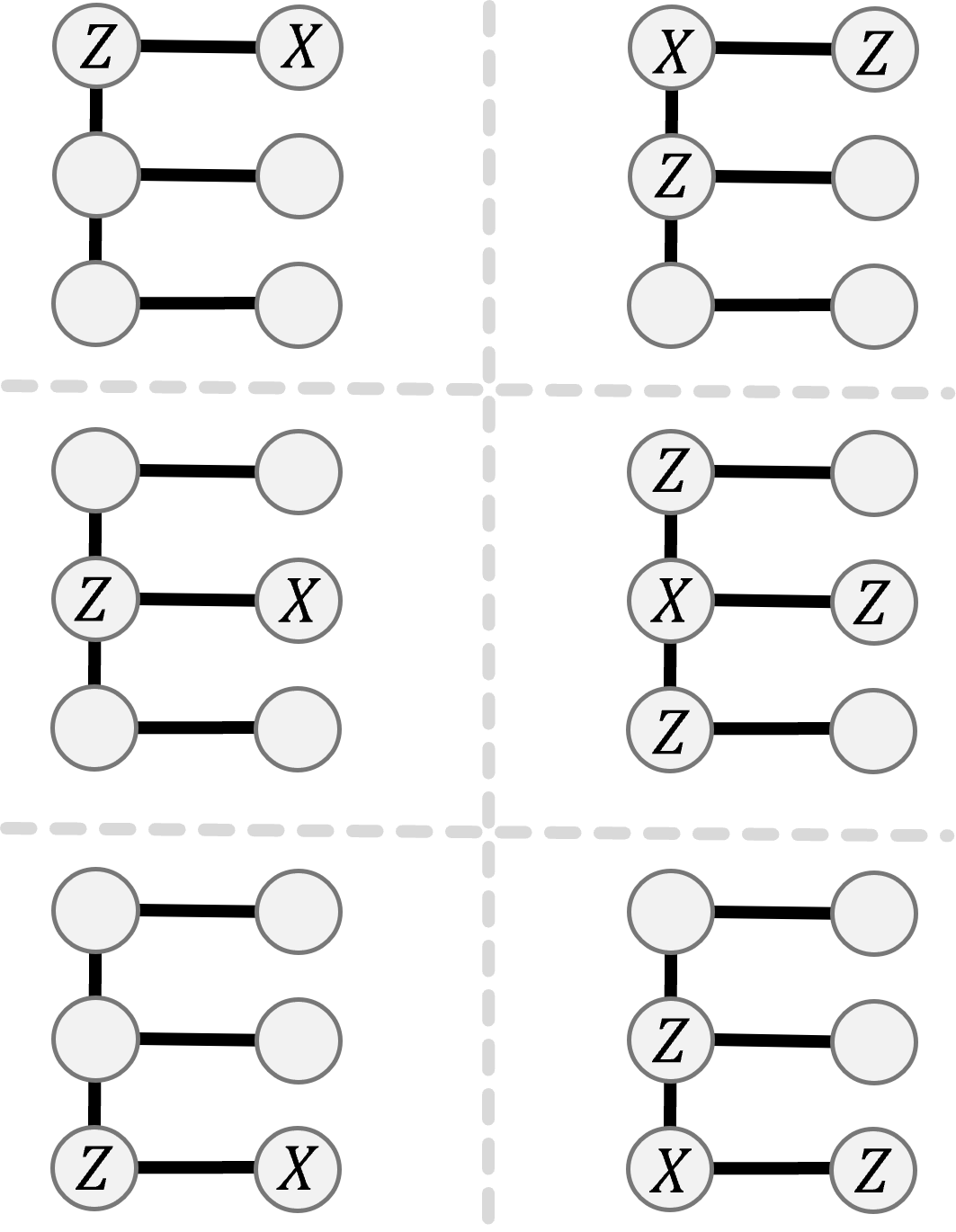} 
\caption{\textbf{Example of a graph state stabilizer.} Shown are the six canonical generators of the stabilizer group (as defined by Eq.~\eqref{eq:graph_stab}) for one specific 6-qubit graph state. They are not unique and can be multiplied in a way that yields six different generators.}
\label{fig:graph_state}
\end{figure}

This is a stabilizer state that can be described by one generator for each vertex as follows,
\begin{align}
    \mathcal{S}_G=\left\langle X_v\prod_{i\in\mathcal{N}_G(v)}Z_i\right\rangle_{v\in V},\label{eq:graph_stab}
\end{align}
where $\mathcal{N}_G(v)=\{v'\in V\textrm{ such that } \{ v, v'\}\in E\}$ is the neighborhood of $v$ given the graph $G$ (see Fig.~\ref{fig:graph_state}). This stabilizer group can be deduced from the circuit that implements $\ket{G}$ in Eq.~\eqref{eq:graph_state}. This is because prior to applying the controlled-$Z$ operators in Eq.~\eqref{eq:graph_state}, we have a stabilizer state given by Eq.~\eqref{eq:x-state} (the all $\ket{+}$-state), which corresponds to the empty graph, i.e., a graph without edges. Since $CZ$s are Clifford circuits, they map products of Pauli operators on products of Pauli operators and, therefore, 
\begin{align*}
    \mathcal{S}_G=U_{\mathcal{S}_G} \mathcal{S}_XU_{\mathcal{S}_G}^\dagger.
\end{align*}
where
\begin{align}
    U_{\mathcal{S}_G}=\prod_{\{ i,j\}\in E}CZ_{i,j}.\label{eq:graph_unitary}
\end{align}
The exact form of $\mathcal{S}_G$ in Eq.~\eqref{eq:graph_stab} then arises naturally from the commutation relation of $CZ_{i,j}$ with $X_i$, as follows,
\begin{align*}
    CZ_{i,j}X_i=X_iZ_jCZ_{i,j}.
\end{align*}
Note that the generators of a stabilizer group are not unique.
%
%

\subsection{Encoding information in stabilizer states}\label{sec:stab_enc}
In the previous subsection we have learned that stabilizer states are special states that have an efficient classical description in terms of a stabilizer group $\mathcal{S}$ that uniquely specifies an $N$-qubit state if $\mathcal{S}$ is generated by $N$ independent stabilizers in $\mathcal{P}_N$. 
This changes if we remove one (or more) stabilizer generators, say $S_i\in\mathcal{S}$, from the group, such that $\mathcal{S}'=\langle S_j\rangle_{j=1,...,i-1,i+1,...,N}$ defines a new stabilizer group. If we do this, then, the projector in Eq.~\eqref{eq:stab_proj} spans a two-dimensional single-qubit subspace which can be defined by its Bloch representation
\begin{align*}
    \ket{\mathcal{S}'(\vec{r})}&\bra{\mathcal{S}'(\vec{r})}\nonumber\\
    &=\frac{1}{2}(\mathbbm{I}+r_1 \bar{X}_i + r_2 \bar{Y}_i + r_3 \bar{Z}_i)\otimes \ket{\mathcal{S}'}\bra{\mathcal{S}'}
\end{align*}
for $\vec{r}\in\mathbbm{R^3}$ and $|\vec{r}|=1$ when assuming pure states.
$\ket{\mathcal{S}'(\vec{r})}$ describes a single-qubit space, the so-called \emph{codespace}, encoded within an $N$-qubit state space. 
Here, we identified the basis of an effectively two-dimensional space of Hermitian matrices with $\bar{X}_i,\bar{Y}_i,\bar{Z}_i\in \mathfrak{su}(2)$ by their commutation relations,
\begin{align*}
    \bar{X}_i&\equiv S_i\\
    \bar{Z}_i&\equiv P\in\mathcal{P}_N\textrm{ s.t. }\{S_i,P\}=0 \wedge [P,S]=0\:\forall S\in\mathcal{S}'\\
    \bar{Y}_i&\equiv i\bar{X}_i\bar{Z}_i,
\end{align*}
where we abbreviated ``such that'' as ``s.t.''.
Note that the assignment of $S_i$ to $\bar{X}_i$ was arbitrary and that these operators are not uniquely defined as they may differ by stabilizers. To indicate this, we often write them as equivalence classes under multiplication by stabilizers, i.e., $[\bar{X}_i]=\{S_i\}/\mathcal{S}'$. These operators that span a qubit subspace are referred to as logical Pauli operators as they encode logical information in terms of a Bloch representation.
In this way, the stabilizer group $\mathcal{S}'$ and the group of logical operators  defines a \emph{stabilizer code}.

Let us illustrate this encoding of qubits through stabilizers by means of an example. Consider two physical qubits where the first qubit encodes one qubit worth of information and the other is an ancilla, initialized in the $\ket{+}$-state, i.e., 
\begin{align}
    \ket{\mathcal{S}(\alpha,\beta)}=(\alpha\ket{0}+\beta\ket{1})\otimes\ket{+}.\label{eq:in_2q_graph_state}  
\end{align}
The corresponding codespace is spanned by one stabilizer and one set of anticommuting Pauli observables,
\begin{align}
    \mathcal{S}&=\langle X_2\rangle\nonumber\\
    [\bar{X}] &= \{ X_1\}/\mathcal{S}\nonumber\\
    [\bar{Z}] &= \{ Z_1\}/\mathcal{S}\label{eq:in_2q_graph}
\end{align}
and we omit (here and in the following) $\bar{Y}$ as it is specified by the product $\bar{X}\bar{Z}$.

Now consider the graph state stabilizer in Eq.~\eqref{eq:graph_stab} for a two-qubit graph $G$ with $V=\{1,2\}, E=\{\{ 1,2\}\}$ and the corresponding unitary $U_{\mathcal{S}_G}=CZ_{1,2}$. We can use this unitary to encode the above state into a so-called graph code yielding,
\begin{align}
        \ket{\mathcal{S}_G'(\alpha,\beta)}=\alpha\ket{0+}+\beta\ket{1-}.\label{eq:enc_2q_graph_state}
\end{align}
Since $U_{\mathcal{S}_G}$ is Clifford, we can immediately specify the new codespace by its basis of observables and its stabilizers as follows,
\begin{align}
    \mathcal{S}_G'&=\langle Z_1X_2\rangle\nonumber\\
    [\bar{X}']&=\{ X_1Z_2\}/\mathcal{S}_G'\nonumber\\
    [\bar{Z}']&=\{ Z_1\}/\mathcal{S}_G'.\label{eq:enc_2q_graph}
\end{align}
We see that the stabilizer $X_1Z_2\in\mathcal{S}_G$ from Eq.~\eqref{eq:graph_stab} now corresponds to a logical operator in $[\bar{X}']$.

\subsection{Measurements on stabilizer states}\label{sec:stab_measure}
%
\begin{figure*}[ht!]\vspace{-0.0cm}
\centering
  \centering
  \includegraphics[width=0.7\textwidth]{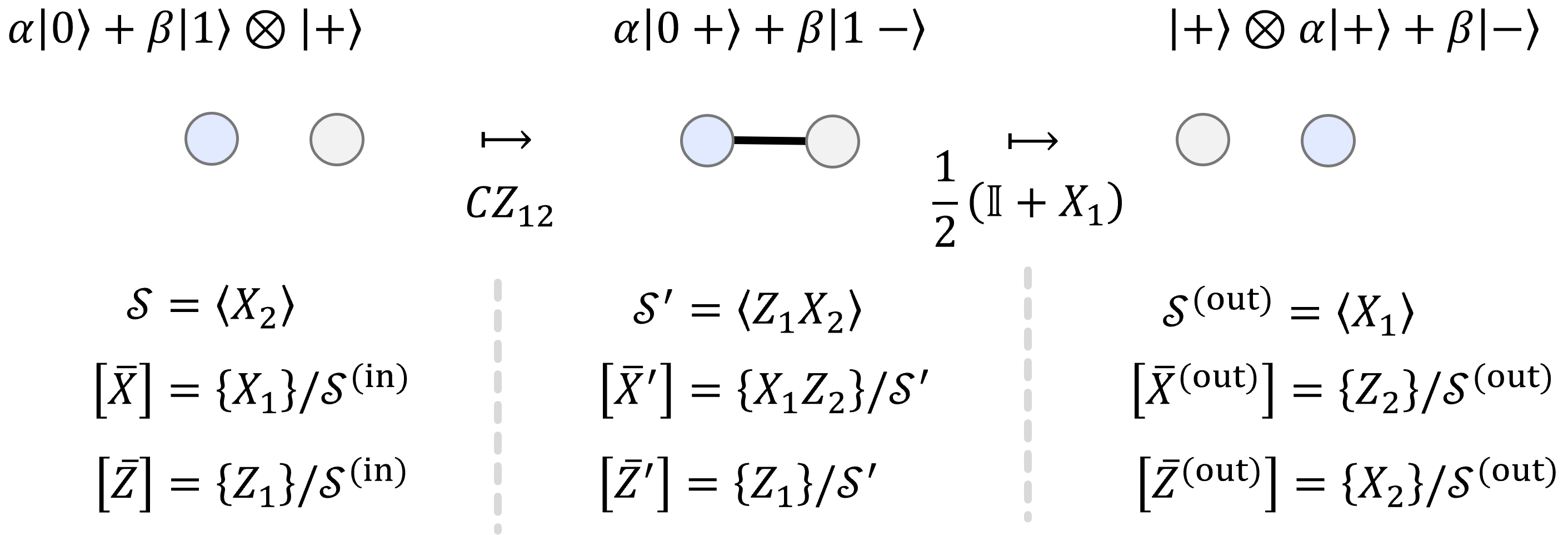} 
\caption{\textbf{Measurement-based teleportation on a 1D graph state.} Teleportation proceeds by initializing an arbitrary state with an ancilla $(\alpha\ket{0}+\beta\ket{1})\otimes \ket{+}$ and entangling them through a controlled-$Z$ gate which corresponds to encoding the input into a graph code. We indicate the specific encoding by coloring the qubit in light blue where the logical information was located before the encoding. Measuring the first qubit (initially encoding the information) in the $X$-basis teleports the information onto the second qubit up to a Hadamard and correction operator (not shown because we assume a projection onto $\ket{+}\bra{+}=\frac{1}{2}(\mathbbm{I}+X)$ for simplicity). The corresponding transformation of the encoded state space (or \emph{codespace}) can either be understood in the Schrödinger picture (top), or in the Heisenberg picture (bottom) where the codespace is defined by the corresponding logical Pauli operators $\bar{X},\bar{Z}$ and stabilizers $S\in\mathcal{S}$. In both pictures, we can infer that the teleportation protocol acts as a logical Hadamard $H$ operation on the codespace for which $HXH=Z$.}
\label{fig:stab_teleport}
\end{figure*}
%
In the previous subsection we have learned that we can lift the constraints given by some stabilizers to obtain logical Pauli operators spanning an observable subspace that allows us to encode qubits into stabilizer codes.
Now, we want to see how we can manipulate the stabilizer and logical subspaces by means of measurements. Measuring a Pauli observable $O\in\mathcal{P}_N$ on a stabilizer state given by $\mathcal{S}$ projects onto a $\pm 1$ eigenstate of $O$ and therefore, modifies the stabilizers as follows,
\begin{align*}
    \mathcal{S}\mapsto \mathcal{S}^{(\textrm{out})}=\{S\in\mathcal{S}\textrm{ s.t. }[S,O]=0\}\times \langle \pm O\rangle,
\end{align*}
where $\times$ indicates multiplication between any pair of elements.
This means that $\pm O$ is added as a stabilizer (where the factor $\pm 1$ indicates different measurement outcomes given by the two possible projections $\frac{1}{2}(\mathbbm{I}\pm O)$) and all stabilizers in $\mathcal{S}$ that do not commute with $O$ must be discarded (by the definition of stabilizers). 
The same holds for the set of logical Pauli observables, i.e.,
\begin{align*}
    [\bar{P}]\mapsto [\bar{P}]^{(\textrm{out})}=\{P\in[\bar{P}]/\mathcal{S}\textrm{ s.t. }[P,O]=0\}/\mathcal{S}^{(\textrm{out})}.
\end{align*}
That is, only those logical Pauli operators that commute with the measured observable remain and are defined up to the modified stabilizer group $\mathcal{S}^{(\textrm{out})}$.

To illustrate the effect of a measurement on the state space defined by a stabilizer group and a set of logical observables, let us consider the graph state example in Eq.~\eqref{eq:enc_2q_graph} (see Fig.~\ref{fig:stab_teleport}). 
As a start, consider a measurement in the $X_1$-basis, i.e., $\frac{1}{2}(\mathbbm{I}\pm X_1)=\ket{\pm}\bra{\pm}$. Without loss of generality we assume a positive measurement outcome and therefore add the stabilizer $X_1$ which transforms Eq.~\eqref{eq:enc_2q_graph} as follows,
\begin{align}
    \mathcal{S}_G'&\mapsto \mathcal{S}_G^{(\textrm{out})}\equiv\langle X_1\rangle\nonumber\\
    [\bar{X}']&\mapsto [\bar{X}^{(\textrm{out})}]\equiv\{Z_2\}/\mathcal{S}_G^{(\textrm{out})}\nonumber\\
    [\bar{Z}_1']&\mapsto[\bar{Z}^{(\textrm{out})}]\equiv\{ X_2\}/\mathcal{S}_G^{(\textrm{out})}.\label{eq:measure_2q_graph}
\end{align}
We see that the information is now encoded in the second physical qubit.

Now we can see that Eqs.~\eqref{eq:in_2q_graph},~\eqref{eq:enc_2q_graph} and~\eqref{eq:measure_2q_graph} describe a teleportation protocol where we have initialized a qubit and ancilla according to Eq.~\eqref{eq:in_2q_graph}, then applied $U_{\mathcal{S}_G}$ to encode the information as in Eq.~\eqref{eq:enc_2q_graph} and finally measured $X_1$ to yield Eq.~\eqref{eq:measure_2q_graph} (see Fig.~\ref{fig:stab_teleport}). This protocol maps the single-qubit logical subspace spanned by $X_1$ and $Z_1$ to a logical single-qubit subspace spanned by $Z_2$ and $X_2$, respectively. That is, we have indeed teleported the logical information up to a Hadamard transformation $H$ which is defined by $HXH=Z$ and $HZH=X$. Note that a negative measurement outcome would have to be treated as an additional $Z_1$ before the measurement due to $Z\ket{+}=\ket{-}$. Given the stabilizer group $\mathcal{S}'_G$ in Eq.~\eqref{eq:enc_2q_graph}, a $Z_1$ corresponds to an $X_2=Z_1\times Z_1X_2$ on the second qubit. This matches the correction operator in the standard teleportation scenario with graph states.

We can easily check our results against the standard state picture. After initializing an input state and ancilla as in Eq.~\eqref{eq:in_2q_graph_state} and encoding it into a graph state as in Eq.~\eqref{eq:enc_2q_graph_state}, we can project the state of qubit 1 into $\ket{\pm}$ such that,
\begin{align*}
    \ket{\mathcal{S}^{(\mathrm{out})}_\pm(\alpha,\beta)}=\ket{\pm}_1\otimes (\alpha \ket{+}_2\pm \beta \ket{-}_2).
\end{align*}
Indeed, we see that $\ket{\mathcal{S}^{(\mathrm{out})}_{\pm}(\alpha,\beta)}$ corresponds to the state space given by Eq.~\eqref{eq:measure_2q_graph}, while the correction, in case of a $\ket{-}$-outcome, is an $X_2$-operator.

If, instead of measuring in the $X_1$-basis, we had measured in the $X_2$- or $Z_1$-basis, we would have effectively measured $X_2,Z_1\in[\bar{Z}_1']$ in Eq.~\eqref{eq:enc_2q_graph} such that the resulting state would have been a stabilizer state given by $\mathcal{S}_G^{(\textrm{out})}=
\langle Z_1, X_2\rangle$.

%
\begin{figure*}[ht!]
\begin{minipage}{\linewidth}
\begin{algorithm}[H]
\begin{algorithmic}
\Require $2N$-qubits, depth $D$, CQCA $T$, $U_T$ (defined in Eq.~\eqref{eq:UT_unitary}), rotation angles ${\vec\theta}=\{\theta_1^1,..., \theta_N^D\}$.\vspace{0.2cm}
\State $j \gets 1$
\State $\ket{\textrm{in}}\gets(\ket{+}^{(\textrm{in})})^{\otimes N}$
\While{$j\leq D$}
\State $\textrm{(1) Add output ancillas:}\:\ket{\textrm{out}}\gets\ket{\textrm{in}}\otimes (\ket{+}^{(\textrm{out})})^{\otimes N}$\vspace{0.2cm}

\State $\textrm{(2) Apply $U_T^{(\textrm{in,out})}$:}\: \ket{\textrm{out}} = U_T^{(\textrm{in,out})} \ket{\textrm{out}}$\vspace{0.2cm}

\State $\textrm{(3) Apply rotations:}\: \ket{\textrm{out}}= \prod_{i=1,...,N}\exp\left(i\theta_i^jZ_i^{(\textrm{in})}\right)\otimes \mathbbm{I}^{(\textrm{out})}\ket{\textrm{out}}$

\State $\textrm{(4) Measure and trace input:}\: \ket{\textrm{out}}= \sqrt{2}^N\prod_{i=1,...,N}\bra{+}_i^{(\textrm{in})}\left(Z_i^{(\textrm{in})}\right)^{m_i^j}\ket{\textrm{out}}$

\State $\textrm{(5) Correct for measurement outcomes }\{m_i^j\}_{i=1,...N}\in\{0,1\}^N:\:\ket{\textrm{out}}=\prod_{i=1,...,N}\left(T(Z_i^{(\textrm{out})})\right)^{m_i^j}\ket{\textrm{out}}$\vspace{0.2cm}

\State $\ket{\textrm{in}}\gets\ket{\textrm{out}}$\vspace{0.2cm}

\State $j=j+1$\vspace{0.2cm}
\EndWhile
\end{algorithmic}
\caption{MBQC by unit cells}
\label{alg:mbqc}
\end{algorithm}
\end{minipage}
\end{figure*}
%
\section{MBQC from CAQC}\label{sec:mbqc}

In this section, we construct an MBQC that implements a CAQC based on an arbitrary CQCA $T$. Since a CAQC consists of unit cells of computation (i.e., $Z$-rotations followed by $T$), we will similarly construct unit cells of MBQC. This formulation will heavily rely on the stabilizer formalism introduced in Sec.~\ref{sec:stabilizers}.

\subsection{Units cells of MBQC.}
\begin{figure*}[ht!]\vspace{-0.0cm}
\centering
  \centering
  \includegraphics[width=0.95\textwidth]{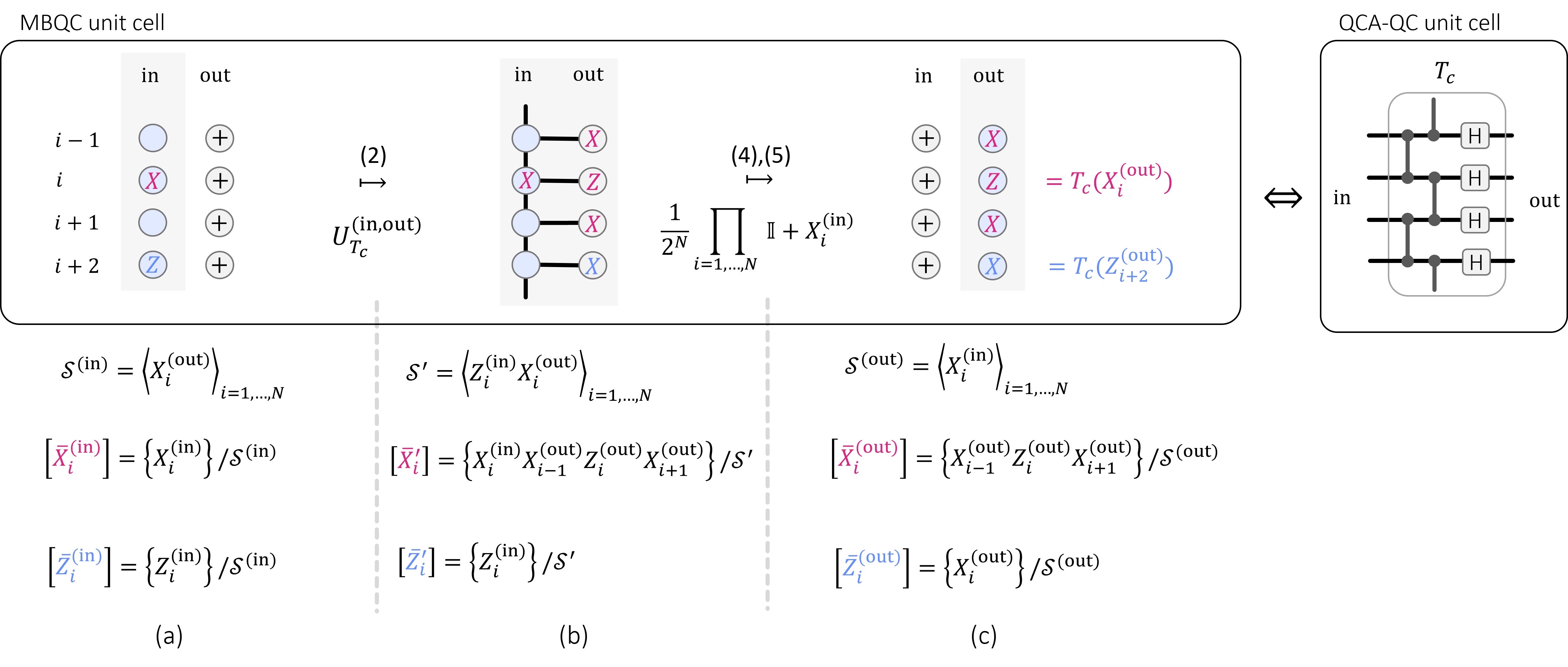} 
\caption{\textbf{Unit cells of MBQC correspond to unit cells of CAQC.} On the left, we show part of a single application of the while-loop from Algorithm~\ref{alg:mbqc} for $T=T_c$ (from Eq.~\eqref{eq:qca_cluster}). After the initialization step~\ref{step:1} (a), $U^{\textrm{in,out}}_{T_c}$ is applied in step~\ref{step:2} according to Eq.~\eqref{eq:UT_unitary} to encode the input into a graph code (b). We indicate the specific encoding by coloring the qubits in light blue where the logical information was located before the encoding. In steps~\ref{step:4} and~\ref{step:5} the input qubits are measured and $\ket{-}$-outcomes are corrected, respectively. In this way, these steps effectively correspond to projections onto the $\ket{+}$-states (c). As the evolution of observables (a)$\rightarrow$(c) suggests, this corresponds to a CAQC unit cell based on $T_c$ without rotations (right).
As we will see in Sec.~\ref{sec:mbqc_stab} and Fig.~\ref{fig:mbqc_stab}, $U_{T_c}$ generates a graph state on a regular square lattice -- that is, a cluster state~\cite{briegel_persistent_2001,raussendorf_oneway_2001} -- and therefore, Algorithm~\ref{alg:mbqc} with $T=T_c$ corresponds to standard MBQC.}
\label{fig:mbqc_from_qcaqc}
\end{figure*}

Consider two parallel chains of $N$ qubits, each defined on a ring (see Fig.~\ref{fig:mbqc_from_qcaqc}). One chain is labeled input $(\textrm{in})$, the other output $(\textrm{out})$. Let $T$ be such that it can act independently on each chain. We define the following (Clifford) unitary $U^{(\textrm{in,out})}_T$ by its action on the basis of observables:

\begin{alignat}{3}
        &U_T^{(\textrm{in,out})}:\nonumber\\[0.2cm] & X_i^{(\textrm{in})}\mapsto X_i^{(\textrm{in})}\otimes T(X_i^{(\textrm{out})}), \hspace{0.6cm} && Z_i^{(\textrm{in})}\mapsto Z_i^{(\textrm{in})}\nonumber\\
        & X_i^{(\textrm{out})}\mapsto Z_i^{(\textrm{in})}\otimes T(Z_i^{(\textrm{out})}), && Z_i^{(\textrm{out})}\mapsto T(X_i^\textrm{out})
        \label{eq:UT}
\end{alignat}
$\forall i=1,...,N$.
We can directly construct this unitary from $CZ$-gates, Hadamards $H$ and $T$ as follows,
\begin{align}\label{eq:UT_unitary}
    U_T^{(\textrm{in,out})}=T^{(\textrm{out})}\prod_{i=1,...,N}H_i^{(\textrm{out})}CZ^{(\textrm{in,out})}_{i,i},
\end{align}
where $CZ^{(\textrm{in,out})}_{i,i}$ indicates that the control qubit is the $i^{\textrm{th}}$ qubit in the $(\textrm{in})$-column and the target is the $i^{\textrm{th}}$ qubit in the $(\textrm{out})$-column.

With this, Algorithm~\ref{alg:mbqc} defines an MBQC on unit cells. In Fig.~\ref{fig:mbqc_from_qcaqc} this algorithm is depicted for $T_c$ which corresponds to an MBQC on the cluster state~\cite{raussendorf_oneway_2001}.

%
%
\subsection{Algorithm~\ref{alg:mbqc} implements a CAQC.} 
Let us understand why this unit cell corresponds to the CAQC as shown in Fig.~\ref{fig:mbqc_from_qcaqc} for the example $T=T_c$. 
Specifically, we will show that one application of the while-loop in Algorithm~\ref{alg:mbqc} maps the basis of the observables $\{ X_i,Z_i\}_{i=1,...,N}$ on the input to the observables $\{ T(X_i),T(Z_i)\}_{i=1,...,N}$ on the output:

\begin{enumerate}[label=(\arabic*)]\setcounter{enumi}{-1}
    \item \label{step:0} Analogous to our discussion in Sec.~\ref{sec:stab_enc}, we consider the basis of observables on the input as follows,
    \begin{align}
        \mathcal{S}^{(\textrm{in})}&=\{\mathbbm{I}^{(\textrm{in})}\}\nonumber\\
        [\bar{X}^{(\textrm{in})}_i]&=\{X^{(\textrm{in})}_i\}\nonumber\\
        [\bar{Z}^{(\textrm{in})}_i]&=\{Z^{(\textrm{in})}_i\}.\label{eq:input_alg}
    \end{align}
   \item \label{step:1} When adding the $\ket{+}^{(\textrm{out})}$-ancillas in the first step, we find that the observables on the input are complemented by a stabilizer state $\mathcal{S}^{(\textrm{in})}=\langle X^{(\textrm{out})}_i\rangle_{i=1,...,N}$ on the output (cf. Fig.~\ref{fig:mbqc_from_qcaqc}(a)). 

    \item \label{step:2} The unitary $U_{T}$ is defined by its action on the basis of observables in Eq.~\eqref{eq:UT}, such that we find the following transformations of the operators in step~\ref{step:1} (cf. Fig.~\ref{fig:mbqc_from_qcaqc}(b)):
    \begin{align*}
        \mathcal{S'}&=\langle Z^{(\textrm{in})}_i \otimes T(Z^{(\textrm{out})}_i)\rangle_{i=1,...,N}\\
        [\bar{X}'_i]&=\{X_i^{(\textrm{in})}\otimes T(X_{i}^{(\textrm{out})})\}/\mathcal{S'}\\
        [\bar{Z}'_i]&=\{Z_i^{(\textrm{in})}\}/\mathcal{S'}.
    \end{align*}
    Note that these are defined up to stabilizers $\mathcal{S'}$ and therefore, $T(Z_i^{(\textrm{out})})\in [\bar{Z}'_i]$. 

\item \label{step:3} From the action of $U_T^{(\textrm{in,out})}$ on $Z_i^{(\textrm{in})}$ in Eq.~\eqref{eq:UT}, we can deduce that $U_T^{(\textrm{in,out})}$ commutes with all $Z$-rotations. So we can absorb the rotations into the input state for now and just understand the action of the algorithm without rotations.

\item \label{step:4} 
Measuring all input qubits in the $\ket{\pm}^{(\textrm{in})}$-basis will randomly project into the state $\ket{+}$ or $\ket{-}$ where $\ket{-}=Z\ket{+}$. Labeling these measurement outcomes by $m_i^j=0,1$, respectively, $\forall i=1,...,N$, we find that measurements formally correspond to some random $Z$-operators $\prod_{i=1,...,N}\left(Z_i^{(\textrm{in})}\right)^{m^j_i}$ applied to the state in step~\ref{step:3} before a projection $\ket{+}\bra{+}$ on all input states. Since $Z_i^{(\textrm{in})}\otimes T(Z_i^{(\textrm{out})})\in\mathcal{S}'$, $\forall i=1,...,N$, we can equivalently write $\left(T(Z_i^{(\textrm{in})})\right)^{m^j_i}$ for the action of the byproduct operators. Keeping this in mind, we find that projecting the input into the $\ket{+}$-state transforms the state space (up to the byproducts on the output and according to our discussion in Sec.~\ref{sec:stab_measure}) as follows  (cf. Fig.~\ref{fig:mbqc_from_qcaqc}(c)),
\begin{align}
  \mathcal{S}^{(\textrm{out})}&=\{\mathbbm{I}^{(\textrm{out})}\}\nonumber\\
    [\bar{X}^{(\textrm{out})}_i]&=\{T(X_i^{(\textrm{out})})\}\nonumber\\
    [\bar{Z}^{(\textrm{out})}_i]&=\{T(Z_i^{(\textrm{out})})\},\label{eq:output_alg}
\end{align}
where we already traced out the input qubits (which are all in the $\ket{+}$-state and, thus, would have corresponded to $X_i^{(\textrm{in})}$-stabilizers)
and we used $T(Z_i^{(\textrm{out})})\in [\bar{Z}'_i]$ in the last equation.

\item \label{step:5} As we have seen in step~\ref{step:4}, the byproduct operators $Z_i^{m_i^j}$ on the input correspond to $T(Z_i^{(\textrm{out})})^{m^j_i}$ on the output which can be directly corrected by applying $T(Z_i^{(\textrm{out})})$ whenever $m_i^j=1$. The necessity to correct for measurement outcomes $m_i^j=1$ within the while-loop, naturally gives rise to a temporal order in MBQC. 
\end{enumerate}

Comparing Eq.~\eqref{eq:input_alg} to Eq.~\eqref{eq:output_alg}, we find that one application of the while-loop in Algorithm~\ref{alg:mbqc} (neglecting rotations for now) acts deterministically as
\begin{align*}
X_i^{(\mathrm{in})}&\mapsto T(X_i^{(\mathrm{out})})\\
Z_i^{(\mathrm{in})}&\mapsto T(Z_i^{(\mathrm{out})})
\end{align*}
for all $i=1,...,N$ which indeed corresponds to one block of CAQC with $T$.

Now consider the rotations that were previously absorbed into the input state. Since we have shown that a single while-loop of Algorithm~\ref{alg:mbqc} acts as $T$ on the full space of observables spanned by $\{X_i,Z_i\}_{i=1,...,N}$ on the input, it also applies to the rotations:
\begin{align*}
    \exp(i\theta X_i^{(\mathrm{in})})&\mapsto\exp(i\theta T(X_i^{(\mathrm{out})}))\\
    \exp(i\theta Z_i^{(\mathrm{in})})&\mapsto\exp(i\theta T(Z_i^{(\mathrm{out})}))
\end{align*}
That is, both the input state and the rotations are transformed according to the local transition rule of the CQCA by one application of the while-loop.
In this way, we indeed recover CAQC from Algorithm~\ref{alg:mbqc} so long as we correct for incorrect measurement outcomes in step~\ref{step:5}.

To conclude, let us formalize the results of this section. Since Algorithm~\ref{alg:mbqc} effectively realizes a CAQC as defined by Theorem~\ref{theorem:qcaqc_gates}, we can directly state the following theorem.
\begin{theorem}\label{theorem:mbqc_unitary}
Consider Algorithm~\ref{alg:mbqc} with depth $D$ acting on two parallel rings of $N$ qubits, using rotation angles $\{\theta_1^1,...,\theta_N^D\}$ and an $N$-qubit CQCA $T$ with a period $L$. Further consider $D$ being a multiple of $L$. This implements the following unitary:
\begin{align}\label{eq:mbqc_unitary}
    \prod_{j=D,...,1}\prod_{i=N,...,1}\exp\left({i\theta^{j}_iT^{[L-[j]_L+1]_L}(Z_i)}\right)
\end{align}
where $[a]_L:=a\mod L$.
\end{theorem}

Note that in standard MBQC~\cite{raussendorf_oneway_2001}, the rotations in step~\ref{step:3} are never performed directly on the state, but are part of the measurement. Clearly, we can also absorb the rotations into the measurements to define a new measurement basis $\ket{\pm_{\theta}}:=\exp(i\theta Z)\ket{\pm}$. Furthermore, standard MBQC is typically not done by unit cells. Instead, a $N\times (D+1)$ resource state is first prepared and then measured. As we will see in Sec.~\ref{sec:mbqc_stab}, we can also define such a resource state for arbitrary CQCAs $T$.

Interestingly, in the same way we can track rotations in a CAQC, we can also track the byproducts to describe an MBQC without corrections (e.g., where all measurements are performed in parallel).
\begin{corollary}\label{corollary:mbqc_uncorrected}
Consider Algorithm~\ref{alg:mbqc} as in Theorem~\ref{theorem:mbqc_unitary} but without correcting measurement results $m_i^j\in\{0,1\}$ in step~\ref{step:5}. This implements the following unitary:
\begin{align}
    \prod_{j=D,...,1}\prod_{i=N,...,1}&\left(T^{[L-[j]_L+1]_L}(Z_i)\right)^{m_i^j}\nonumber\\
    &\times\exp\left({i\theta^{j}_iT^{[L-[j]_L+1]_L}(Z_i)}\right)\label{eq:mbqc_uncorrected}
\end{align}
\end{corollary}
Note that Eq.~\eqref{eq:mbqc_uncorrected} can be further simplified by commuting the byproducts through the whole circuit to the very end using the commutation relation $\exp(i\theta Z)X=X\exp(-i\theta Z)$. In this way, we can understand byproducts as a Pauli circuit at the end of the computation given by $\left(T^{[L-[j]_L+1]_L}(Z_i)\right)^{m_i^j}$ as well as some rotations where the sign of their measurement angles has been flipped $\theta\mapsto -\theta$. Using a description of CQCAs in terms of Laurent polynomials, as described for example in Refs.~\cite{schlingemann_on_2008,stephen_subsystem_2019}, the sign changes can be tracked efficiently classically.

\subsection{Constructing stabilizer states from CQCAs}\label{sec:mbqc_stab}
Instead of performing the MBQC by unit cells as in Algorithm~\ref{alg:mbqc}, we can also create the $N \times (D+1)$ resource state on a square lattice by applying $U_T^{(i,i+1)}$ $D$-times column-by-column on  $N \times (D+1)$ $\ket{+}$-states and then perform the rotated measurements and corrections column-wise.
The resulting unitary is the same as in Theorem~\ref{theorem:mbqc_unitary} simply because $U_T^{(i,i+1)}$ does not act on the qubits at depths $j\leq i-1$ which implies that measurements of column $i$  have to be made after applying $U_T^{(j,j+1)}$ $\forall j=1,..,i$ but not necessarily before $U_T^{(l,l+1)}$ for $l=i+1,...,D$.
The main difference is that correction operators in step~\ref{step:5} may change because $U_T^{(i+1,i+2)}$ may act nontrivially on the byproducts arising from measurements of column $i$.

\begin{figure}[ht!]\vspace{-0.0cm}
\centering
  \centering
  \includegraphics[width=0.9\linewidth]{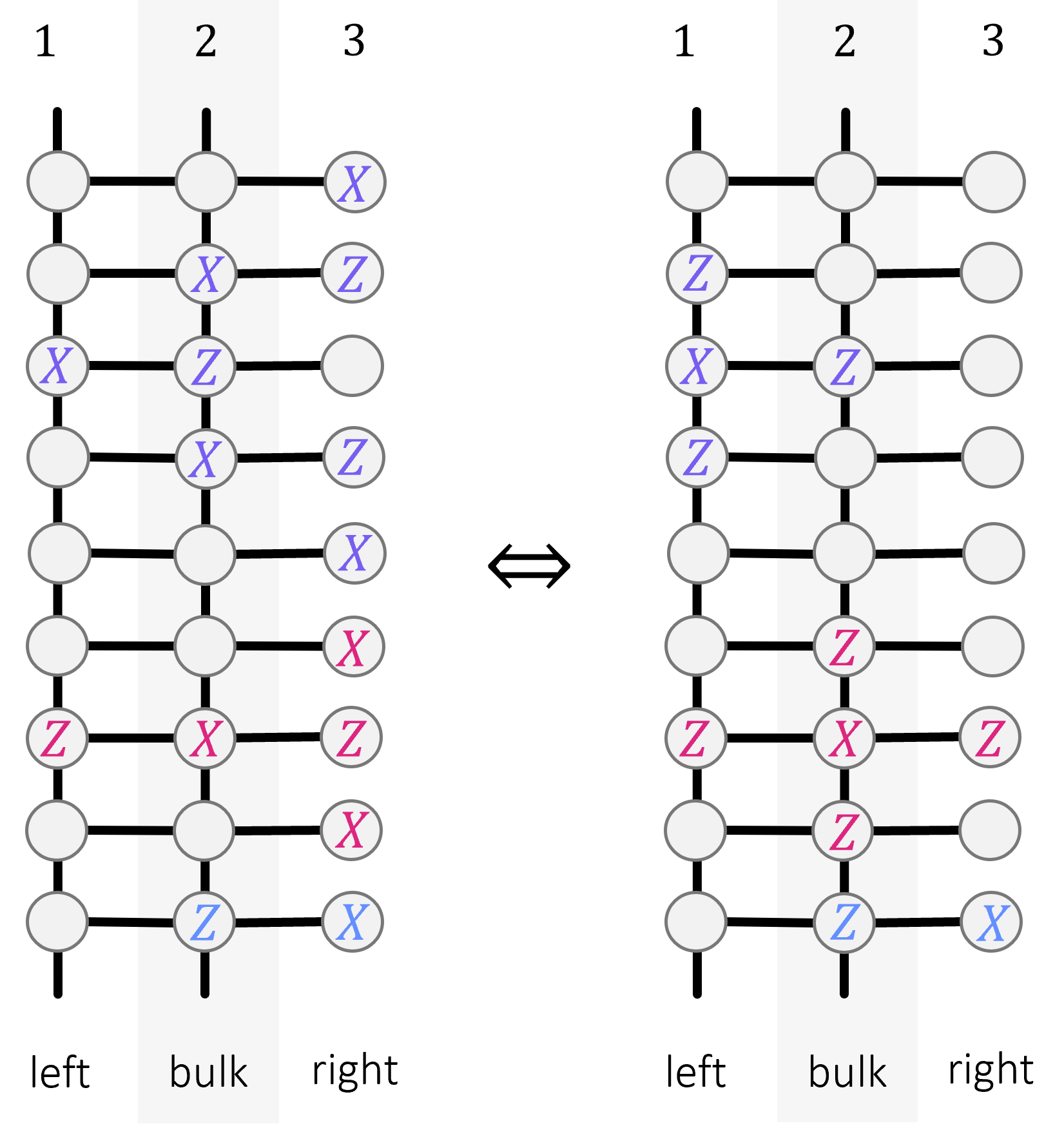} 
\caption{\textbf{Constructing the MBQC resource state from $U_{T_c}$.} 
(left) From top to bottom we depict the stabilizers that correspond to Eq.~\eqref{eq:ut_stab_boundary} (purple), Eq.~\eqref{eq:ut_stab} (magenta) and Eq.~\eqref{eq:ut_stab_right} (blue), respectively, for $T=T_c$ from Eq.~\eqref{eq:qca_cluster}. These stabilizers are defined for every row $i=1,...,N$ and therefore form a generating set of nonlocal stabilizers. By multiplying them together as described in the main text, we obtain an equivalent set of local generators (right) which corresponds to the standard graph state stabilizers of the depicted graph. To generate the full resource state, we keep adding columns and applying $U_{T_c}$ which extends the bulk and moves the boundary to the right. But since $U^{(j,j+1)}_{T_c}$ for $j\geq 3$ acts trivially on the purple and magenta stabilizers, they remain stabilizer generators of the resource state.}
\label{fig:mbqc_stab}
\end{figure}

Since $T$ is a CQCA, Eq.~\eqref{eq:UT} implies that $U_T$ is a Clifford circuit and, therefore, creates a stabilizer state when applied to the initial stabilizer state $\ket{+}^{N\times (D+1)}$ (see Sec.~\ref{sec:stabilizers}). In the following, we will construct the generators of this stabilizer group from $U_T$. 
Importantly, as $U_T$ is locality preserving, the generators are \emph{local}. That means that they have nontrivial support within a constant-size region (i.e., a region that does not scale in size with $N$ or $D$) of the square lattice in which we laid out qubits (see for example Fig.~\ref{fig:mbqc_stab}).

To construct the stabilizers, consider three columns of $N$-qubit rings where we apply $U_T$ between columns $1,2$ and between columns $2,3$, i.e., $U_T^{(2,3)}U_T^{(1,2)}$. The initial $X_i^{(j)}$-generators (i.e., stabilizers of the $\ket{+}$-states) are transformed as follows (cf. Eq.~\eqref{eq:UT}),
\begin{align}
    X_i^{(1)}&\mapsto X_i^{(1)}\otimes T(X_i^{(2)})\otimes \prod_{k\in \mathcal{X}(T(X_0))}T(X^{(3)}_{i+k}),\label{eq:ut_stab_boundary}\\
    X_i^{(2)}&\mapsto Z_i^{(1)}\otimes T(Z_i^{(2)})\otimes \prod_{k\in \mathcal{X}(T(Z_0))}T(X^{(3)}_{i+k})
    \nonumber\\
    &=Z_i^{(1)}\otimes T(Z_i^{(2)})\otimes T^2(Z^{(3)}_i)\prod_{k\in \mathcal{Z}(T(Z_0))}T(Z^{(3)}_{i+k}),\label{eq:ut_stab}\\
    X_i^{(3)}&\mapsto Z_i^{(2)}\otimes T(Z_i^{(3)}),\label{eq:ut_stab_right}
\end{align}
where we ignored potential phases $\mu\in U(1)$ that may appear in Eqs.~\eqref{eq:ut_stab_boundary},~\eqref{eq:ut_stab} and $\mathcal{X}(T(\cdot))\subset\{1,...,N\}$ is the set of qubits on which $T(\cdot)$ acts as $X$ or $Y$. Such terms appear in this way because $U_T^{(2,3)}$ leaves $Z_i^{(2)}$ invariant for any $i$ and only transforms $X_i^{(2)},Y_i^{(2)}$ nontrivially. In Eq.~\eqref{eq:ut_stab}, we have expressed this term as transformations of $Z$ through the relation
\begin{align*}
    T^2(Z_i)=\nu\prod_{k'\in \mathcal{X}(T(Z_0))}T(X_{i+k'})\prod_{k\in \mathcal{Z}(T(Z_0))}T(Z_{i+k})
\end{align*}
where $\mathcal{Z}(T(\cdot))$ is the set of qubits on which $T(\cdot)$ acts as $Z$ or $Y$ and $\nu\in U(1)$ is another phase which we will also ignore in the following. 

Due to the translational invariance (along rows) of the state construction described above, these equations are sufficient to define the left and right boundary (i.e., leftmost and rightmost column) as well as bulk stabilizers for an arbitrary sized lattice, respectively (see e.g., Fig.~\ref{fig:mbqc_stab}).
To see this, let us start by reducing Eq.~\eqref{eq:ut_stab} to the local stabilizer of the bulk. To this end, we need to introduce the following Lemma adapted from Ref.~\cite{stephen_subsystem_2019},
\begin{lemma}\label{lemma:cqca_t2}
    Consider a CQCA with transition function $T$ acting on a ring of $N$ qubits. Then there exist a constant $m$ independent of N and $\alpha_1,...,\alpha_m, \beta\in\{0,1\}$ such that,
    \begin{align}
        T^2(Z_i)=\omega Z_i\prod_{k=1,...,m}\biggl(T(Z_{i-k})T(Z_{i+k})\biggr)^{\alpha_k}T(Z_i)^\beta\label{eq:lemma_t2}.
    \end{align}
    where $\omega\in U(1)$ is an arbitrary phase.
\end{lemma}
Note that a similar identity as Eq.~\eqref{eq:lemma_t2} but with different $\alpha_k,\beta$ also holds for $T^2(X_i)$ (which we will not use here). Also, as we have done before, we will ignore the phases $\omega$ in the following.

Lemma~\ref{lemma:cqca_t2} now allows us to rewrite $T^2(Z_i^{(3)})$ in Eq.~\eqref{eq:ut_stab} only in terms of products of $Z^{(3)}$ and $T(Z^{(3)})$. Then, we can use the stabilizers $Z_i^{(2)}\otimes T(Z_i^{(3)})$ in Eq.~\eqref{eq:ut_stab_right} to further simplify Eq.~\eqref{eq:ut_stab} such that it acts on columns $1,3$ as $Z_i$. That is, the following equation is equivalent to Eq.~\eqref{eq:ut_stab} under multiplication of Eq.~\eqref{eq:ut_stab_right},
\begin{align}
    X_i^{(2)}&\mapsto Z_i^{(1)}\otimes\nonumber\\ &T(Z_i^{(2)})\prod_{k=1,...,m}\left(Z_{i-k}^{(2)}Z_{i+k}^{(2)}\right)^{\alpha_k}\left(Z_i^{(2)}\right)^{\beta}\nonumber\\
    &\times\prod_{k'\in \mathcal{Z}(T(Z_0))}Z^{(2)}_{i+k'}\otimes Z_i^{(3)}
\end{align}
As further columns are added and $U_T$ are applied, this stabilizer remains unchanged because any other $U_T$ acts as identity on columns $1,2$ and $Z_i^{(3)}$ by definition. As the above argument applies not only to column 2, but in general to any column that is not a boundary, we have found a local representation of the bulk stabilizer group.

Let us now identify the left and right boundary stabilizers from Eqs.~\eqref{eq:ut_stab_boundary}-\eqref{eq:ut_stab_right}.
Because we apply $U_T$ from left to right, the transformation rule of the rightmost $X$-stabilizer in Eq.~\eqref{eq:ut_stab_right} already defines a local stabilizer generator for the right boundary at $X_i^{D+1}$.
To identify the leftmost boundary, we find that the local operator $T^{-1}(Z^{(1)}_i)\otimes Z^{(2)}_i$ commutes with all stabilizers in Eqs.~\eqref{eq:ut_stab_boundary}-\eqref{eq:ut_stab_right} because
\begin{align*}
    \bigl(P\otimes T(P)\bigr)&\bigl(T^{-1}(Q)\otimes Q\bigr)\\
    &=P T^{-1}(Q)\otimes T(P)Q\\
    &=T^{-1}\bigl(T(P)Q\bigr)\otimes T(P)Q\\
    &=T^{-1}\bigl(QT(P)\bigr)\otimes QT(P)\\
    &=T^{-1}(Q)P\otimes QT(P)\\
    &=\bigl(T^{-1}(Q)\otimes Q\bigr)\bigl(P\otimes T(P)\bigr)
\end{align*}
which holds for all $P,Q\in\mathcal{P}_N$. 
Since any following $U_T^{(j,j+1)}$ for $j\geq 3$ does not act on $T^{-1}(Z^{(1)}_i)\otimes Z^{(2)}_i$, this stabilizer remains unchanged under further extension of the stabilizer group in our construction and therefore, defines the left boundary condition.
In Fig.~\ref{fig:mbqc_stab}, we exemplify this construction using the cluster CQCA $T_c$ as defined in Eq.~\eqref{eq:qca_cluster} which we will also discuss in detail later.
%
%

Let us now summarize the results of this section in terms of a theorem.
\begin{theorem}\label{theorem:stab_state}
    Consider a CQCA $T$ with $T(Z_i)\neq Z_i$ and $D+1$ rings of $N$ qubits aligned in parallel and initialized in $\ket{+}^{N\times (D+1)}$. Applying $\prod_{j=D,...,1}U_T^{(j,j+1)}$, where $U_T^{(j,j+1)}$ is defined by its action on two neighboring rings $j,j+1$ as in Eq.~\eqref{eq:UT}, gives rise to a stabilizer state given by a stabilizer group $\mathcal{S}_T$ as follows,
    \begin{align}
        \mathcal{S}_T&= \Big{\langle}S^{(1)}_i,S^{(j+1)}_i,S^{(D+1)}_i\Big{\rangle}_{\begin{subarray}{l}j=1,...,D-1\\ i=1,...,N\end{subarray}}\nonumber\\&\equiv\Big{\langle} T^{-1}(Z^{(1)}_i)\otimes Z^{(2)}_i,\nonumber\\ 
        &Z_i^{(j)}\otimes T(Z_i^{(j+1)})\nonumber\\
        &\times\prod_{k=1,...,m}\left(Z_{i-k}^{(j+1)}Z_{i+k}^{(j+1)}\right)^{\alpha_k}\left(Z_i^{(j+1)}\right)^{\beta}\nonumber\\
        &\times\prod_{k'\in \mathcal{Z}(T(Z_0))}Z^{(j+1)}_{i+k'}\nonumber\\&\otimes Z_i^{(j+2)},\nonumber\\
        &Z_i^{(D)}\otimes T(Z_i^{(D+1)})\Big{\rangle}_{\begin{subarray}{l}j=1,...,D-1\\ i=1,...,N\end{subarray}}\label{eq:theorem_stab}
    \end{align}
    for some fixed $\alpha_1,...\alpha_m,\beta\in\{0,1\}$ given by Lemma~\ref{lemma:cqca_t2} and up to potential phases $\omega_i^{(j)}\in U(1)$.
\end{theorem}

Note that we assumed $T(Z_i)\neq Z_i$ in Theorem~\ref{theorem:stab_state}. This is because the generators as defined in Eq.~\eqref{eq:theorem_stab} are  independent by construction only if $T(Z_i)\neq Z_i$. If instead $T(Z_i)=Z_i$, we have $T(X_i)=X_iZ_i^{\eta_0}\prod_{k=1,...,m}\left(Z_{i+k}Z_{i-k}\right)^{\eta_k}$ for some constant $m<N$ and $\eta_k\in\{0,1\}$ such that we find the following stabilizer group:
\begin{align*}
    \mathcal{S}_T=\Big\langle Z_i^{(j)}\otimes Z_{i}^{(j+1)}, X_i^{(1)}\bigotimes_{k=2,...,D+1}T(X_i^{(k)})\Big\rangle_{\begin{subarray}{l}j=1,...,D\\ i=1,...,N\end{subarray}}
\end{align*}
If $D$ is even, this corresponds to $N$ $(D+1)$-qubit GHZ-states. If $T$ is entangling and $D$ is odd, these GHZ-states are also entangled. This difference between odd and even $D$ stems from the fact that such a CQCA is periodic. An example of such a CQCA is $\tilde{T}_c$ in Eq.~\eqref{eq:qca_periodic}. In Sec.~\ref{sec:nonsimple}, we will discuss how to recover a universal resource state from such CQCAs.

Let us now consider performing an MBQC on the full stabilizer state defined by Eq.~\eqref{eq:theorem_stab} instead of using Algorithm~\ref{alg:mbqc}. As explained at the beginning of this section, only the byproducts have to be handled differently. Specifically, the intermediate correction operation for a byproduct $Z^{(j)}_i$ arising from a measurement at $i\in\{1,...,N\}$ and $j\in\{1,...,D\}$ is given by the complementing bulk or (right) boundary stabilizer generator $Z^{(j)}_i\:S^{(j+1)}_i$ at position $i,j+1$. In that way, any appearing byproduct can be corrected to become a stabilizer which, by definition, acts trivially on the state space.

As a first example, let us construct the standard MBQC on a cluster state. To this end, consider the CQCA $T_c$ defined in Eq.~\eqref{eq:qca_cluster}. The corresponding unitary $U_{T_c}$ is given by Eqs.~\eqref{eq:UT} and~\eqref{eq:UT_unitary}.
Let us illustrate the construction of the stabilizer group according to Theorem~\ref{theorem:stab_state} (see Fig.~\ref{fig:mbqc_stab}). With $m=1$, $\alpha_1=1,\beta=0$ we can confirm Lemma~\ref{lemma:cqca_t2} as we have $T_c^2(Z_i)=X_{i-1}Z_iX_{i+1}=T_c(Z_{i-1})Z_iT_c(Z_{i+1})$. Since $\mathcal{Z}(T_c(Z_0))=\emptyset$, we find the following stabilizers for the bulk:
\begin{align*}
    \mathcal{S}^{\textrm{bulk}}_{T_c}=\Big{\langle} Z_i^{(j)}X_i^{(j+1)}Z_{i-1}^{(j+1)}Z_{i+1}^{(j+1)}Z_i^{(j+2)}\Big{\rangle}_{\begin{subarray}{l}j=1,...,D-1\\ i=1,...,N\end{subarray}},
\end{align*}
which corresponds to the standard graph state stabilizer group (see Eq.~\eqref{eq:graph_stab}) on a regular square lattice without boundaries.

The boundaries of the lattice are defined at $j=1,D$. Since $T_c^{-1}(Z_i)=Z_{i-1}X_iZ_{i+1}$ and $T_c(Z_i)=X_i$, we find the following boundary stabilizers according to Eq.~\eqref{eq:theorem_stab},
\begin{align*}
    \mathcal{S}_{T_c}^{\textrm{boundary}}=\langle X_i^{(1)}Z_{i-1}^{(1)}Z_{i+1}^{(1)}Z_i^{(2)},
    Z_i^{(D)}X_i^{(D+1)}\rangle_{i=1,...,N},
\end{align*}
which correspond to the standard graph state stabilizers for the boundaries shown in Fig.~\ref{fig:mbqc_stab}.
In this way, $\mathcal{S}^{\textrm{bulk}}_{T_c}\times \mathcal{S}^{\textrm{boundary}}_{T_c}$ defines a cluster state on a regular square lattice with periodic boundary conditions along the top and bottom and input/output boundaries along the left/right. The correspondence between Eqs.~\eqref{eq:ut_stab_boundary}-\eqref{eq:ut_stab_right} and Eq.~\eqref{eq:theorem_stab} is shown in Fig.~\ref{fig:mbqc_stab} for this example.

In general, Eq.~\eqref{eq:theorem_stab} simplifies significantly for simple CQCA $T_s$. Remember that for any simple CQCA, $T_s(X_i)=Z_i\prod_{k=1,...,m}\left(X_{i+k}X_{i-k}\right)^{\eta_k}$ for some constant $m<N$ and $\eta_k\in\{0,1\}$ to preserve the appropriate commutation relations with $T_s(Z_i)=X_i$ for all $i=1,...,N$. Similarly, we can conclude $T^{-1}(Z_i)=X_i\prod_{k=1,...,m}\left(Z_{i+k}Z_{i-k}\right)^{\eta_k}$. By the construction of the stabilizers in Eq.~\eqref{eq:theorem_stab}, we find that they correspond to local stabilizers that are also the stabilizers of a local graph state.
Interestingly, this means that any resource state defined by $\mathcal{S}_{T_s}$ in Eq.~\eqref{eq:theorem_stab} for a simple CQCA $T_s$ can be prepared in constant depth even though our construction implicitly requires a depth linear in $D$.

%
\begin{figure}[t!]\vspace{-0.0cm}
\centering
  \centering
  \includegraphics[width=0.9\linewidth]{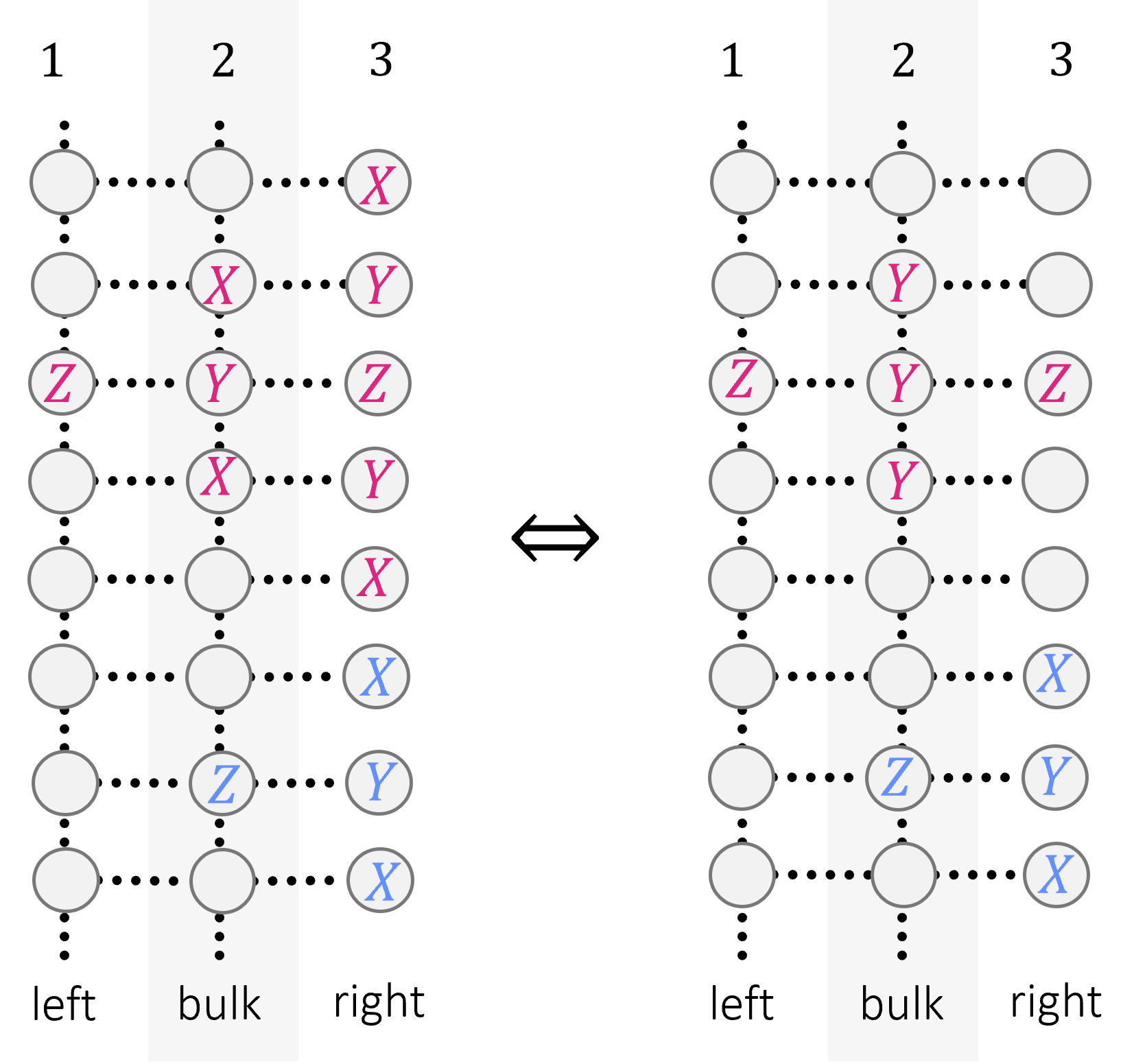} 
\caption{\textbf{Constructing the MBQC resource state from $U_{\hat{T}_c}$.} 
(left) From top to bottom we depict the stabilizers that correspond to Eq.~\eqref{eq:ut_stab} (magenta) and Eq.~\eqref{eq:ut_stab_right} (blue), respectively, for $T=\hat{T}_c$ from Eq.~\eqref{eq:qca_fractal}. These stabilizers are defined for every row $i=1,...,N$ and form a generating set of nonlocal stabilizers together with the left boundary stabilizers $\hat{T}_c^{-1}(Z_i^{(1)})\otimes Z_i^{(2)}$ (not shown). By multiplying them together as described in the main text, we obtain an equivalent set of local generators (right) which corresponds to a non-standard resource state for MBQC and is not a canonical graph state (which is why we do not draw edges here).}
\label{fig:mbqc_stab_fractal}
\end{figure}
%
As a second, nontrivial example consider the CQCA $\hat{T}_c$ in Eq.~\eqref{eq:qca_fractal} (see Fig.~\ref{fig:mbqc_stab_fractal}). 
Let us construct the corresponding stabilizer state from $U_{\hat{T}_c}$ in Eq.~\eqref{eq:UT}. For brevity, we will only consider the bulk stabilizers $\mathcal{S}^{\textrm{bulk}}_{\hat{T}_c}$. In line with Lemma~\ref{lemma:cqca_t2}, $m=1, \alpha_1=1,\beta=1$ such that $\hat{T}_c^2(Z_i)=X_{i-2}Z_{i-1}X_iZ_{i+1}X_{i+2}=-i\hat{T}_c(Z_{i-1})\hat{T}_c(Z_i)\hat{T}_c(Z_{i+1})Z_i$. Since $\mathcal{Z}(\hat{T}_c(Z_0))=\{0\}$, we find, in accordance with Theorem~\ref{theorem:stab_state} and as shown in Fig.~\ref{fig:mbqc_stab_fractal},
\begin{align*}
        \mathcal{S}^{\textrm{bulk}}_{\hat{T}_c}=\Big{\langle} Z_i^{(j)}Y_i^{(j+1)}Y_{i-1}^{(j+1)}Y_{i+1}^{(j+1)}Z_i^{(j+2)}\Big{\rangle}_{\begin{subarray}{l}j=1,...,D-1\\ i=1,...,N\end{subarray}},
\end{align*}
up to some phases. The boundary stabilizers are easily identified given $\hat{T}_c^{-1}(Z_i)=Y_{i-1}X_iY_{i+1}$.

\subsection{Universal MBQC from non-simple CQCAs}\label{sec:nonsimple}
%
\begin{figure}[b!]\vspace{-0.0cm}
\centering
  \centering
  \includegraphics[width=\linewidth]{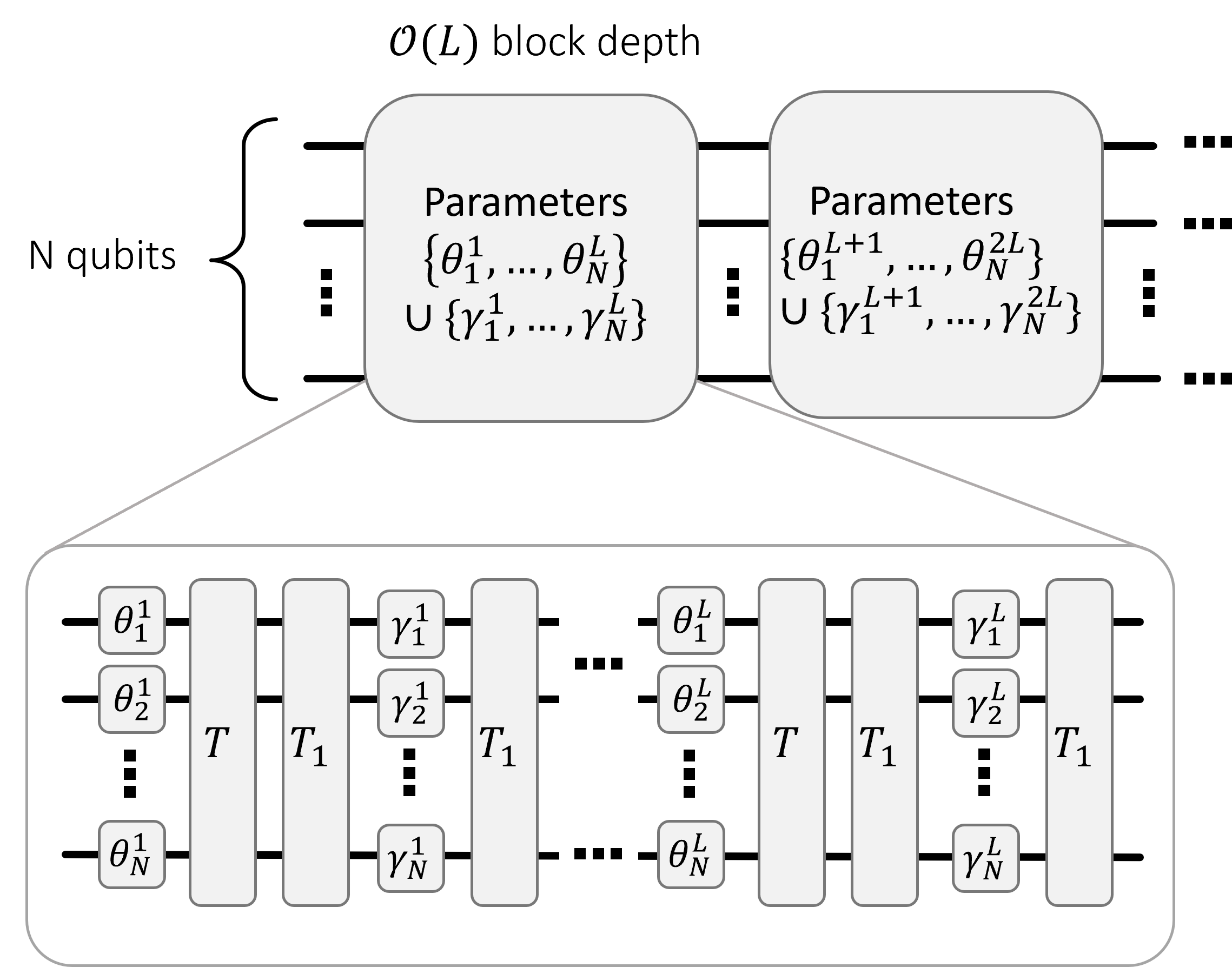} 
\caption{\textbf{Universal CAQC for non-simple CQCA.} Non-simple CQCA $T$ may still be used to generate universal gate sets in an extended CAQC formalism. To this end, we write $T=T_1T'$ where $T'=T_1T$ and $T_1$ is a periodic CQCA defined in Eq.~\eqref{eq:qca_1d} such that $T_1\exp(i\gamma Z_i)T_1=\exp(i\gamma X_i)$. Then, alternating between applications of $T'$ and $T_1$, each preceded by local $Z$-rotations, we naturally obtain a universal gate set within each block if $T$ is entangling (see Eq.~\eqref{eq:qcaqc_extended_gates}.}
\label{fig:extended_qcaqc}
\end{figure}
%
So far, we have restricted ourselves to simple CQCAs to prove universality through Theorems~\ref{theorem:qcaqc_gates},~\ref{theorem:mbqc_unitary}.
Similarly, we can easily prove universality for all entangling CQCA $T$ with period $L$ for which there exist $k\in \{1,...,L\}$ such that $T^k(Z_i)=X_i,Y_i$. This is because the entangling property allows us to construct entangling gates according to Eq.~\eqref{eq:qcaqca_gates} while $T^L(Z_i)$ and $T^k(Z_i)$ give us a universal single-qubit gate.
However, so long as a CQCA is entangling, even if it does not have the above property, it can be extended in a way that gives rise to a universal CAQC as well as a universal resource state in MBQC.
This is done by considering products of CQCAs instead of single CQCAs. That is, instead of applying the same, possibly non-simple, CQCA $T$ every unit cell, we alternate between two different CQCAs $T_1,T_2$ such that $T=T_1T_2$ (see Fig.~\ref{fig:extended_qcaqc}).
Importantly, this preserves the periodicity of the original CQCA as $(T_1T_2)^L=T^L=id$ such that the total depth of each block of the CAQC in Fig.~\ref{fig:extended_qcaqc} is only doubled.

While we can, in principle, choose any such product of CQCAs $T_1T_2$ to extend $T$, we make a specific choice such that any entangling CQCA $T$ extended in this way becomes universal with respect to the unitary implemented by the corresponding CAQC.
That is, we consider the following non-entangling, periodic CQCA, which acts as Hadamards on all qubits,
\begin{align}\label{eq:qca_1d}
    T_1(X_i)&=Z_i\nonumber\\
    T_1(Z_i)&=X_i.
\end{align}
Since $T_1^2=id$, we can write, $T=T_1T_2$ for $T_2=T_1T$. 
Now consider a CAQC as in Fig.~\ref{fig:extended_qcaqc}. Since $T_1(Z_i)=X_i$, each block of CAQC naturally implements a logical $X$-rotation through the last layer of $Z$-rotations. Since $T$ is entangling, there is also an entangling gate available and $Z$-rotations come for free in the first layer due to $T^L=id$. Since the periodicity is preserved, we have proven universality for any entangling CQCA $T$ that is extended by $T_1$.

Similar to Theorem~\ref{theorem:qcaqc_gates}, we can formally present the resulting unitary in the following proposition:
\begin{prop}\label{prop:qcaqc_extended}
    Consider a CQCA $T$ with a period $L$ acting on a ring of $N$ qubits. Let a block of CQCA-based quantum computation consist of $L$ repeated, sequential applications of $T_1T$ followed by $T_1$ (see Eq.~\eqref{eq:qca_1d}), where the $j^{\mathrm{th}}$ application of $T_1T$ is preceded by $N$ $Z$-rotations $\exp(i\theta^j_iZ_i)$ acting on qubits $i=1,...,N$ with parameters $\{\theta_1^j,...,\theta_N^j\}$ and the $j^{\mathrm{th}}$ application of $T_1$ is preceded by $N$ $Z$-rotations $\exp(i\gamma^j_iZ_i)$ acting on qubits $i=1,...,N$ with parameters $\{\gamma_1^j,...,\gamma_N^j\}$  for all $j=1,...,L$ (see Fig.~\ref{fig:extended_qcaqc}). This block implements the following unitary:
\begin{align}\label{eq:qcaqc_extended_gates}
    \prod_{j=L,...,1}\prod_{i=N,...,1}
    &\exp\left({i\gamma^{j}_iT^{L-j}(X_i)}\right)\nonumber\\&\times\exp\left({i\theta^{j}_iT^{L-j+1}(Z_i)}\right).
\end{align}
\end{prop}

Since Algorithm~\ref{alg:mbqc} is implemented column-by-column, we can easily extend it to allow for two (or more) alternating CQCA using unitaries $U_{T_1T}$ and $U_{T_1}$ (see Eq.~\eqref{eq:UT}). In this way, we can construct universal MBQC from non-simple CQCA which implements unitaries of the form Eq.~\eqref{eq:qcaqc_extended_gates}.

\begin{figure}[b!]\vspace{-0.0cm}
\centering
  \centering
  \includegraphics[width=0.75\linewidth]{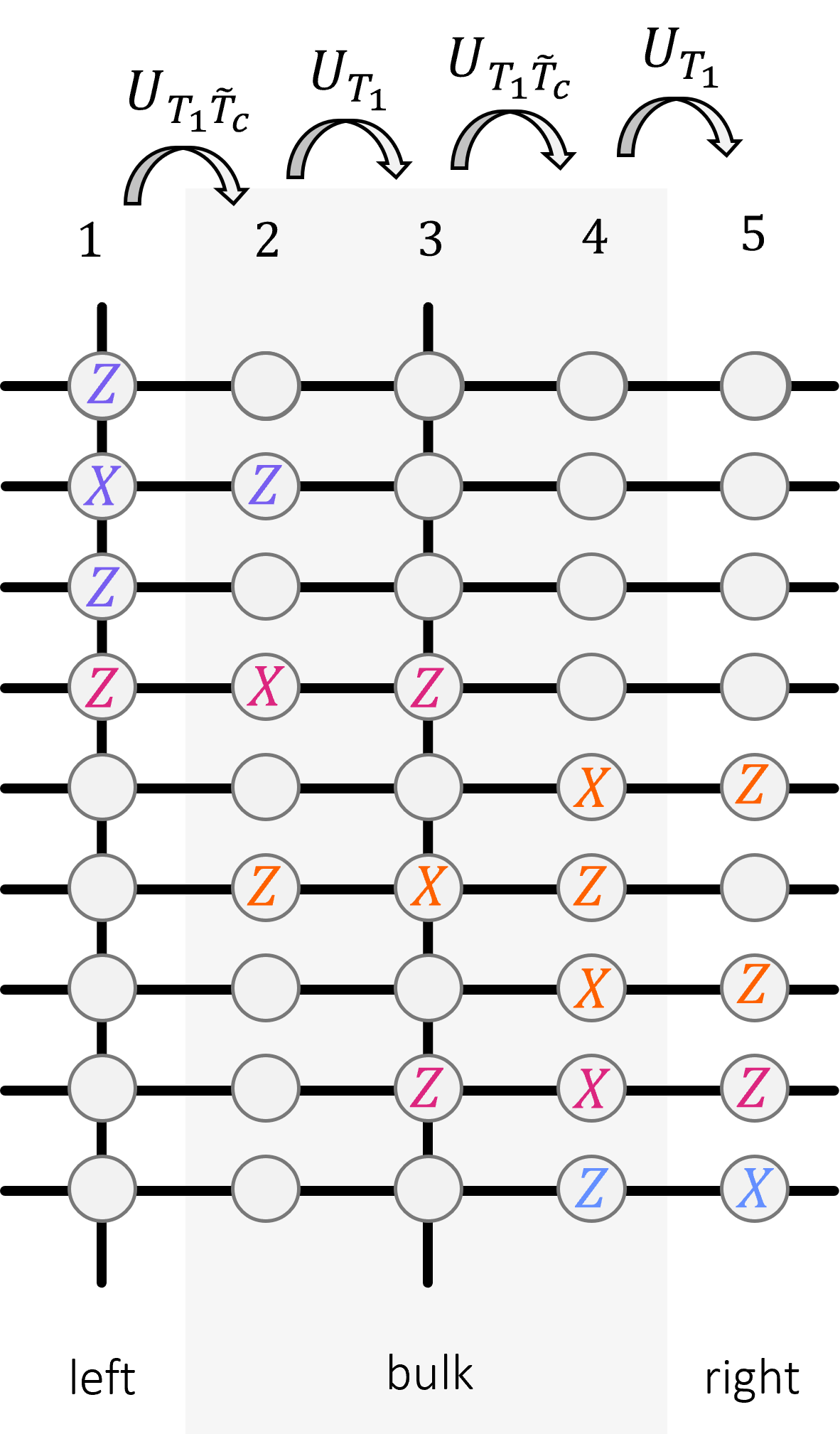} 
\caption{\textbf{Constructing the MBQC resource state from products of $U_{T_1}$ and $U_{T_1\tilde{T}_c}$.} From top to bottom, representatives from the generating set of local stabilizers in Eq.~\eqref{eq:stab_extended} are shown. They are constructed using the transformation rules in Eq.~\eqref{eq:UT} for sequential applications of $U_{T_1}$ and $U_{T_1\tilde{T}_c}$ for $\tilde{T}_c$ from Eq.~\eqref{eq:qca_periodic} and $T_1$ from Eq.~\eqref{eq:qca_1d}. The bulk stabilizers (orange) can be further simplified to obtain the local graph state stabilizers in Eq.~\eqref{eq:periodic_stab_extended}.}
\label{fig:extended_stab}
\end{figure}

Importantly, we can also construct local stabilizer generators for the extended formalism. 
This can be done by considering 5 columns of local stabilizer states $\langle X^{(j)}_i\rangle_{i,j}$ and applying $U^{(4,5)}_{T_1}U^{(3,4)}_{T_1T}U^{(2,3)}_{T_1}U^{(1,2)}_{T_1T}$ according to Eq.~\eqref{eq:UT}. As exemplified in Fig.~\ref{fig:extended_stab}, this immediately yields local stabilizer generators because $U_{T_1}^{(j-1,j)}X^{(j-1)}_i (U_{T_1}^{(j-1,j)})^\dagger=X^{(j-1)}_i\otimes Z^{(j)}_i$ and $U_T^{(j,j+1)}$ leaves $Z^{(j)}_i$ invariant for any $T$. We can summarize this result in the following Proposition:
\begin{prop}\label{prop:stab_extended}
        Consider a non-simple CQCA $T$ and $2D+1$ rings of $N$ qubits aligned in parallel and initialized in $\ket{+}^{N\times (2D+1)}$. Let $T'=T_1T$ with $T_1$ from Eq.~\eqref{eq:qca_1d}.  
        Applying $U_{T_1}^{(2D,2D+1)}\prod_{j=D-1,...,1}\left(U_{T'}^{(2j+1,2j+2)}U_{T_1}^{(2j,2j+1)}\right)U_{T'}^{(1,2)}$, where $U_T^{(j,j+1)}$ is defined by its action on two neighboring rings $j,j+1$ as in Eq.~\eqref{eq:UT}, gives rise to a stabilizer state given by a stabilizer group $\mathcal{S}_T'$ as follows,
    \begin{align}\label{eq:stab_extended}
        \mathcal{S}_T'&= \Big{\langle}S'^{(1)}_i,S'^{(2j)}_i,S'^{(2j+1)}_i,S'^{(2D+1)}_i\Big{\rangle}_{\begin{subarray}{l}j=1,...,D-1\\ i=1,...,N\end{subarray}}\nonumber\\
        &\equiv\Big{\langle} T'^{-1}(Z^{(1)}_i)\otimes Z^{(2)}_i,\nonumber\\ 
        &Z_i^{(2j-1)}\otimes T'(Z_i^{(2j)})\otimes \prod_{k\in\mathcal{X}(T'(Z_0))}Z^{(2j+1)}_{(i+k)},\nonumber\\
        &Z_i^{(2j)}\otimes X_i^{(2j+1)}\otimes T'(X_i^{(2j+2)})\otimes \prod_{k\in\mathcal{X}(T'(X_0))}Z^{(2j+2)}_{i+k},\nonumber\\
        &Z_i^{(2D)}\otimes X_i^{(2D+1)}\Big{\rangle}_{\begin{subarray}{l}j=1,...,D-1\\ i=1,...,N\end{subarray}}
    \end{align}
        up to potential phases $\omega_i^{(l)}\in U(1)$.
\end{prop}

%
%
Note that we have to treat $T(Z_i)=X_i$, i.e., simple CQCAs, separately because in that case $T'(Z_i)=Z_i$ and our construction yields dependent stabilizers. The treatment of this case is analogous to before and omitted here. Note further that the particular choice of local generators in Eq.~\eqref{eq:stab_extended} may be amenable to simplification under multiplication of its elements. However, here our main goal was to provide a generating set of local stabilizers.

Let us consider the example in Fig.~\ref{fig:extended_stab} where we initially started with the periodic CQCA $\tilde{T}_c$ from Eq.~\eqref{eq:qca_periodic} which yields a non-universal resource state in the original construction. In the extended construction this non-simple CQCA yields a universal gate set by Proposition~\ref{prop:qcaqc_extended}. Since $T_c=T_1\tilde{T}_c$, we find the following bulk stabilizers (see Fig.~\ref{fig:extended_stab}):
\begin{align}\label{eq:periodic_stab_extended}
    \mathcal{S}&_{\tilde{T}_c}'^{\:\textrm{bulk}}=\Big\langle Z_i^{(2j-1)}\otimes X_i^{(2j)}\otimes Z_i^{(2j+1)}, \nonumber\\
    &Z_i^{(2j)}\otimes Z_{i-1}^{(2j+1)}X_i^{(2j+1)}Z_{i+1}^{(2j+1)}\otimes Z_{i}^{(2j+2)}\Big\rangle_{\begin{subarray}{l}j=1,...,D-1\\ i=1,...,N\end{subarray}}
\end{align}
which is a \emph{decorated} cluster state with additional qubits on edges along the horizontal. 
Indeed, we find that for any CQCA $T$ for which $T_1T$ is simple our extended construction yields a graph state as in the original construction but decorated with additional qubits along horizontal edges.

\section{MBQC-based Ansätze for PQCs}\label{sec:pqc}
%
\begin{figure*}[ht!]\vspace{-0.0cm}
\centering
  \centering
  \includegraphics[width=0.95\textwidth]{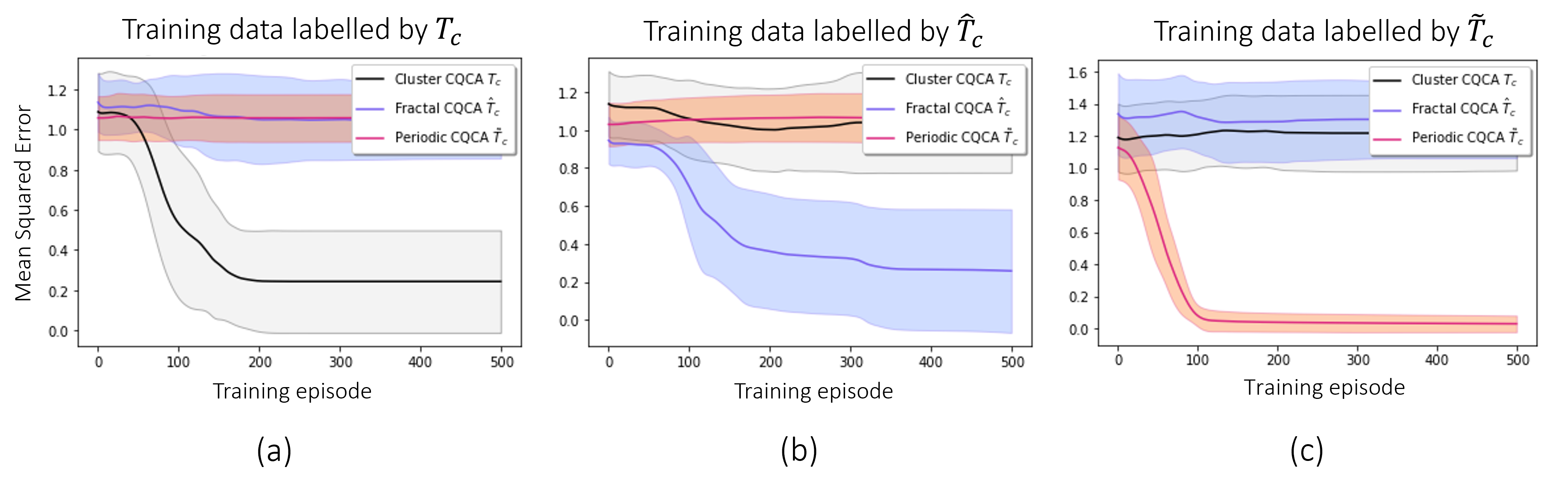} 
\caption{\textbf{Training MBQC-based Ansätze for PQCs on stilted quantum datasets.} We consider PQC models based on CQCAs $T_c$ (see Eq.~\eqref{eq:qca_cluster}) and $\hat{T}_c$ (see Eq.~\eqref{eq:qca_fractal}) to implement the unitaries in Eq.~\eqref{eq:qcaqca_gates} and one model based on CQCA $\tilde{T}_c$ (see  Eq.~\eqref{eq:qca_periodic}) to implement the unitary in  Eq.~\eqref{eq:qcaqc_extended_gates}. Each model is trained to learn labels on a fashion MNIST dataset labeled by each model. All plots are averaged over 10 models with different initializations. (a) All three models are trained on a dataset labeled by the model based on $T_c$. The best $T_c$-based model achieves a loss of $10^{-8}$ while the best other model achieves a loss of $0.8$. (b) All three models are trained on a dataset labeled by the $\hat{T}_c$-based model. The best $\hat{T}_c$-based model achieves a loss of $10^{-8}$ while the best other model achieves a loss of $0.7$. (c) All three models are trained on a dataset labeled by the $\tilde{T}_c$-based model. The best $\tilde{T}_c$-based model achieves a loss of $10^{-8}$ while the best other model achieves a loss of $0.8$.}
\label{fig:pqc_results}
\end{figure*}
%
In this section, we will investigate the unitaries in Eqs.~\eqref{eq:qcaqca_gates} and~\eqref{eq:qcaqc_extended_gates} as Ansätze for PQCs. That is, we use the family of unitaries implementable by an MBQC or CAQC as learning models for quantum machine learning.\footnote{The simulations are available at \href{https://www.github.com/HendrikPN/parameterized-mbqc-ansatz}{github.com/HendrikPN/parameterized-mbqc-ansatz}.}
Specifically, we consider a supervised learning model which, given some classical input vector $\bf{x}\in\mathbbm{R}^N$, learns to associate a label $y(\textbf{x})\in [0,1]$ with it. To this end, the classical input vector is encoded into a quantum circuit using the feature encoding proposed in Ref.~\cite{havlivcek_supervised_2019}. This encoding circuit is followed by a PQC based on CAQC as illustrated in Figs.~\ref{fig:qcaqc_blocks} and~\ref{fig:extended_qcaqc}. At the end, we evaluate the expectation value for a $Z$-observable on qubit $1$ to represent the learned label (which is normalized to obtain a standard deviation of $1$ for all system sizes). 

Our goal is to demonstrate that different CQCA can yield PQC Ansätze with significantly varying performances on varying tasks. To this end, we compare three different PQC Ansätze, two of which are based on Eq.~\eqref{eq:qcaqca_gates} for the cluster CQCA $T_c$ (see Eq.~\eqref{eq:qca_cluster}) and the fractal CQCA $\hat{T}_c$ (see Eq.~\eqref{eq:qca_fractal}), respectively, while a third Ansatz is based on Eq.~\eqref{eq:qcaqc_extended_gates} for the periodic CQCA $\tilde{T}_c$ in Eq.~\eqref{eq:qca_periodic}. The first PQC would then correspond to a standard MBQC on the cluster state (see Fig.~\ref{fig:mbqc_stab}), the second one would correspond to an MBQC on the resource state in Fig.~\ref{fig:mbqc_stab_fractal} and the third PQC would correspond to an MBQC on a horizontally decorated cluster state (see Fig.~\ref{fig:extended_stab}).

Similar to Ref.~\cite{huang_power_2021}, we consider a regression task with input data from the fashion MNIST dataset~\cite{xiao_fashion_2017} composed of $28\times 28$-pixel images of clothes. Using principal component analysis, we first reduce the
dimension of these images to obtain $N$-dimensional input vectors and then label the images through our learning models with a random set of parameters. That is, we consider a total of three learning tasks where each of our three learning models is trained on a dataset that has been labeled by the others and itself, which is also called a \emph{stilted} quantum dataset. 

The results are presented in Fig.~\ref{fig:pqc_results} for $N=6$ qubits and depth $D=4$. We can see that each model is able to learn only if the labels have been produced by the same model which suggests that the family of learning models based on CAQC (or equivalently, MBQC) yields a broad and diverse pool of Ansätze for PQCs. This is particular interesting for the design of problem-specific (instead of hardware-efficient, or universal) Ansätze which are relevant to avoid barren plateaus.
While not shown in Fig.~\ref{fig:pqc_results}, it is noteworthy that the learnability degrades for all models with increasing $N$ and $D$, which is most likely due to the presence of barren plateaus~\cite{larocca_diagnosing_2022,goh_liealgebraic_2023} or absence of an in-depth hyperparameter analysis.

\section{Conclusion}\label{sec:conclusion}
In this work, we have described the popular measurement-based model of computation in terms of a CQCA-based quantum computation (which we summarized in Theorem~\ref{theorem:qcaqc_gates}). We have shown that, given a CQCA $T$, we can construct an MBQC through Algorithm~\ref{alg:mbqc} that implements a set of unitaries according to Theorem~\ref{theorem:mbqc_unitary}. In this way, MBQC can be described by a CAQC in the circuit model where we can even track byproduct operators using Corollary~\ref{corollary:mbqc_uncorrected}. 
We have further shown that the corresponding resource states for MBQC are local stabilizer states according to Theorem~\ref{theorem:stab_state}.

We have further shown that MBQC based on a CQCA $T$ (as described by Theorem~\ref{theorem:mbqc_unitary}) is universal if $T$ is simple and entangling. Then, the stabilizer group of the corresponding resource state (as described in Theorem~\ref{theorem:stab_state}) simplifies to local graph states.
Through Propositions~\ref{prop:qcaqc_extended} and~\ref{prop:stab_extended}, we have then shown that any non-universal MBQC (or CAQC) can be extended to a universal MBQC (or CAQC).

The relation between CQCAs and MBQC that was discussed in this work has given rise to an interesting connection between MBQC and certain topological phases of matter, called symmetry-protected topological phases~\cite{raussendorf_computationally_2019,stephen_subsystem_2019}, through tensor networks. Namely, there exists a notion of computational phases of matter where computational universality as it arises from Theorem~\ref{theorem:mbqc_unitary} persists throughout a family of states. That is, given a stabilizer state $\ket{\mathcal{S}_T}$ as in Theorem~\ref{theorem:stab_state}, we can define a symmetry-protected topological phase as a parameterized family of states containing $\ket{\mathcal{S}_T}$ such that measurements in the $XY$-plane yield universal quantum computation.

Moreover, we have discussed the usefulness of CAQC (or MBQC) as an Ansatz for parameterized quantum circuits (PQCs). Here, different CQCAs and combinations thereof can be used to not only yield hardware-efficient but also problem specific Ansätze. The latter approach is particularly relevant to avoid barren plateaus~\cite{larocca_diagnosing_2022,goh_liealgebraic_2023}.
Interestingly, the family of PQCs that arises from CQCAs connects well to the mathematical framework of Refs.~\cite{larocca_diagnosing_2022,goh_liealgebraic_2023} due to the simple form of the generating Lie algebra for the unitaries in Theorem~\ref{theorem:qcaqc_gates}.
Moreover, it was shown in Ref.~\cite{majumbder_2023_variational}, that parameterized MBQC is a beneficial approach to generative modeling because partially
uncorrected measurements can be exploited to add controlled randomness to the learning model using Corollary~\ref{corollary:mbqc_uncorrected}. 

Both CAQC and MBQC are particularly suitable for highly parallelizable architectures such as neutral atoms controlled by optical tweezers~\cite{bluvstein_quantum_2022}. Such architectures enable an efficient implementation of the translationally invariant operations needed to create the resource state in MBQC or for implementing the CQCA in CAQC.

\section*{Acknowledgements}
We would like to thank I.D.~Smith, D.~Orsucci and S.~Jerbi for useful discussions. This research was funded in whole, or in part, by the Austrian Science Fund (FWF) DOI 10.55776/F71. For open access purposes, the authors have applied a CC BY public copyright license to any author-accepted manuscript version arising from this submission.
We gratefully acknowledge support by the European Union (ERC, QuantAI, Project No. $101055129$). Views and opinions expressed are however those of the author(s) only and do not necessarily reflect those of the European Union or the European Research Council. Neither the European Union nor the granting authority can be held responsible for them.




\bibliographystyle{stdWithTitle}
\bibliography{bib}

\end{document}